\documentclass[%
 aip,
 amsmath,amssymb,
 reprint,%
floatfix]{revtex4-1}

\usepackage{graphicx}
\usepackage{dcolumn}
\usepackage{bm}
\usepackage[binary-units]{siunitx}
\usepackage{enumitem}
\usepackage{mathptmx, physics}
\usepackage{color}
\usepackage[usenames,dvipsnames]{xcolor}
\usepackage[colorlinks=true,linkcolor=Maroon,citecolor=OliveGreen,urlcolor=Blue,linktoc=page]{hyperref}
\pdfoutput=1

\usepackage{tikz}

\usepackage{easyReview}

\definecolor{myblue}{HTML}{0000CD}
\definecolor{myorange}{HTML}{568203}
\definecolor{mypurple}{HTML}{800080}

\renewcommand{\vec}[1]{\boldsymbol{#1}}
\newcommand{\Tc}{T_{c}}
\newcommand{\Jc}{J_{c}}
\newcommand{\Jdp}{J_{d}}
\newcommand{\kB}{k_{\mathrm{B}}}
\newcommand{\Hctwo}{H_{c2}}
\newcommand{\Hcone}{H_{c1}}
\newcommand{\Hc}{B_{c}}

\newcommand{\circled}[1]{\tikz[baseline=(char.base)]{\node[shape=circle,draw,fill=lightgray,inner sep=0.7pt] (char) {#1};}}

\begin{document}

\preprint{AIP/123-QED}

\title{Perspective: Challenges and Transformative Opportunities in Superconductor Vortex Physics}

\author{Serena Eley}
\email{serenaeley@mines.edu}
\affiliation{
Department of Physics, Colorado School of Mines, Golden, Colorado, USA
}

\author{Andreas Glatz}%
\affiliation{
Materials Science Division, Argonne National Laboratory, 9700~S.~Cass~Avenue,~Argonne,~Illinois~60639,~USA
}
\affiliation{
Department of Physics, Northern Illinois University, DeKalb, Illinois 60115, USA
}

\author{Roland Willa}%
\affiliation{
Theory of Condensed Matter Physics,~Karlsruhe Institute of Technology,~Karlsruhe,~Germany 
}%
\affiliation{
Heidelberger Akademie der Wissenschaften, Heidelberg, Germany 
}%

\date{\today}

\begin{abstract}

In superconductors, the motion of vortices introduces unwanted dissipation that is disruptive to applications. Fortunately, material defects can immobilize vortices, acting as vortex pinning centers, which engenders dramatic improvements in superconductor material properties and device operation. This has motivated decades of research into developing methods of tailoring the disorder landscape in superconductors to increase the strength of vortex pinning. Yet efficacious materials engineering still alludes us.  The electromagnetic properties of real (disordered) superconducting materials cannot yet be reliably predicted, such that designing superconductors for applications remains a largely inefficient process of trial and error. This is ultimately due to large gaps in our knowledge of vortex dynamics: the field is challenged by the extremely complex interplay between vortex elasticity, vortex-vortex interactions, and material disorder.  


In this Perspective, we review obstacles and recent successes in understanding and controlling vortex dynamics in superconducting materials and devices. We further identify major open questions and discuss opportunities for transformative research in the field.  This includes improving our understanding of vortex creep, determining and reaching the ceiling for the critical current, advanced microscopy to garner accurate structure-property relationships, frontiers in predictive simulations and the benefits of artificial intelligence, as well as controlling and exploiting vortices in quantum information applications.

\end{abstract}

\maketitle

\section{\label{sec:Introduction}Introduction}

Distinguished for their ability to carry high dissipation-less currents below a critical temperature $\Tc$, superconductors are used in motors, generators, fault-current limiters, and particle accelerator magnets. Their impact spans beyond these examples of large-scale applications, also affecting nanoscale devices. Perhaps most renown for their key role in the quantum revolution, superconductors constitute building blocks in current and next-generation devices for computing and sensing.  For example, superconducting photons detectors feature high-resolutions due to high kinetic inductance and a sharp superconductor-to-normal phase transition. Moreover, superconductors can be configured to form anharmonic oscillators that can be exploited in quantum computing. 

\begin{figure}[ht]
\includegraphics[width = 0.4\textwidth]{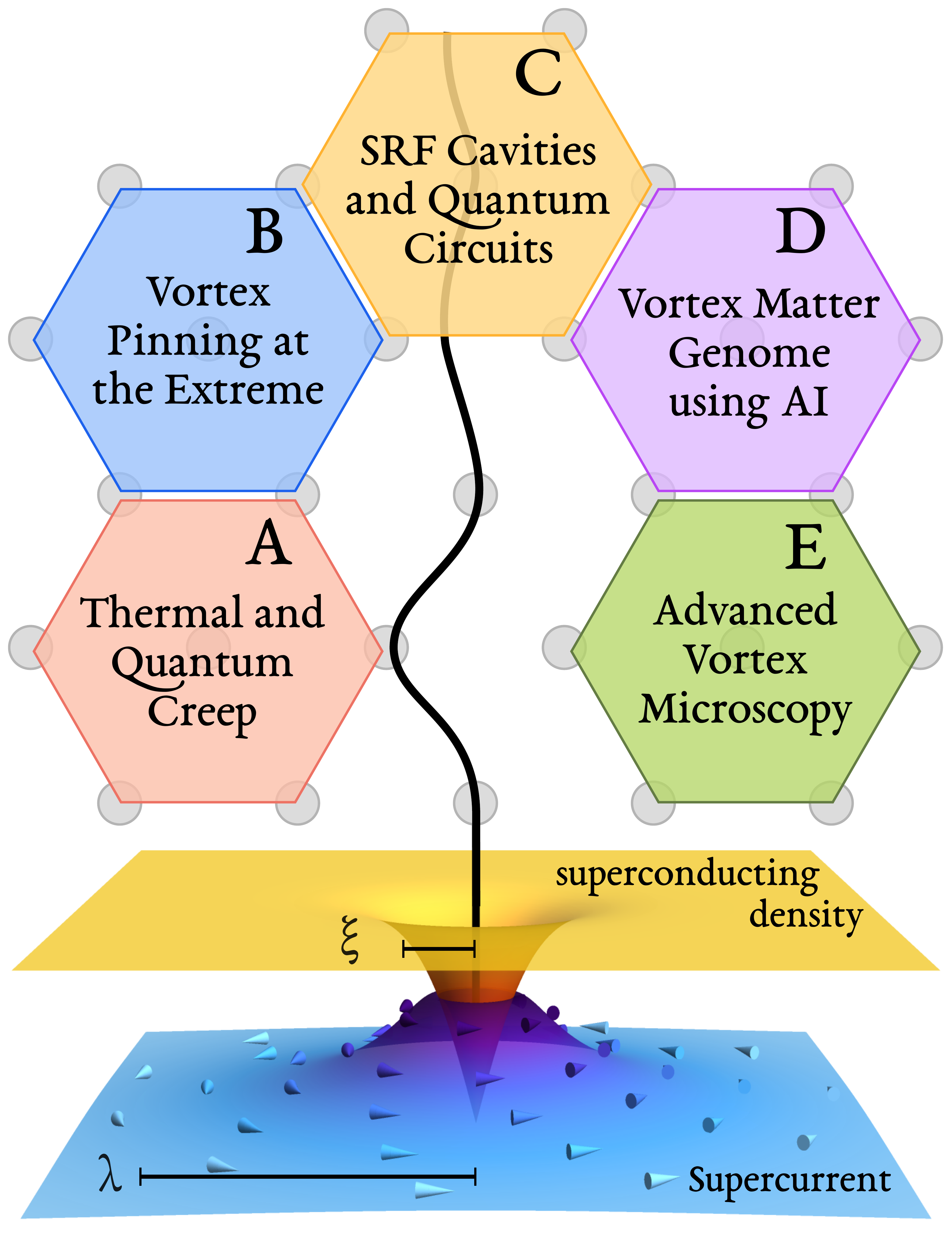}
\caption{Frontiers in vortex matter research. Black line represents the vortex core. Yellow region shows how the density of superconducting electron pairs decays towards the center of the core (of size $\sim \xi$, coherence length). Blue plane (with arrows) represents amplitude of supercurrent, circulating around the core of radius up to the penetration depth $\lambda$.
\label{fig:summary}
}
\end{figure}
\begin{figure}[htp]
\includegraphics[width=1\linewidth]{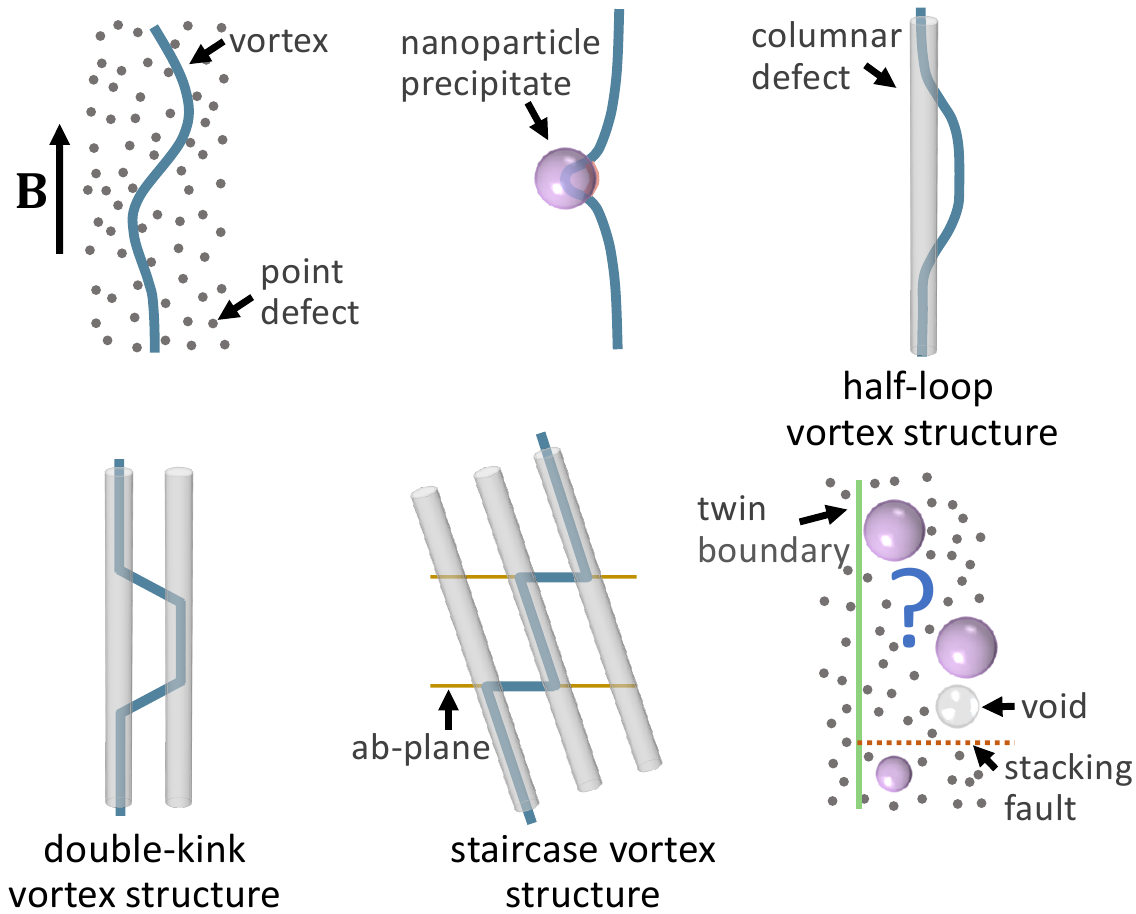}
\caption{\label{fig:vortexstructures} Examples of vortex structures (curved blue lines) that are predicted to form in different defect landscapes under the influence of an applied current. Imaging these structures and defects, would allow us to establish the crucial connection between vortex excitations, vortex-defect and vortex-vortex interactions, Lorentz forces, and resulting vortex phases that is needed for efficacious defect engineering.}
\end{figure}

Notwithstanding these successes, the performance of superconducting devices is often impaired by the motion of vortices---lines threading a quantum $\Phi_{0} = h / 2e$ of magnetic flux through the material (see Fig.~\ref{fig:summary}). Propelled by electrical currents and thermal/quantum fluctuations, vortex motion is dissipative such that it limits the current-carrying capacity in wires, causes losses in microwave circuits, contributes to decoherence in qubits, and can also induce phase transitions.  Understanding vortex dynamics is a formidable challenge because of the complex interplay between moving vortices, material disorder that can counteract (pin) vortex motion, and thermal energy that causes vortices to escape from these pinning sites. Furthermore, as depicted in Fig.~\ref{fig:vortexstructures}, in three-dimensional samples (bulk crystals or thick films), vortices are elastic objects that form complicated shapes as they wind through the disorder landscape, reshaping and moving under the influence of current-induced Lorentz forces.

These complexities encumber predictability: we can neither predict technologically important parameters in superconductors nor prescribe an ideal defect landscape that optimizes these parameters for specific applications.  Though modifying the disorder landscape, e.g. using particle-irradiation or by incorporating non-superconducting inclusions into growth precursors, can engender dramatic enhancements in the current carrying capacity, these processes are often designed through a trial-and-error approach.  Furthermore, the optimal defect landscape is highly-material dependent. This is because the efficacy of pinning centers depends on the relationship between their geometry and the vortex structure, the latter being determined by parameters of the superconductor such as the coherence length $\xi$, penetration depth $\lambda$, and the anisotropy $\gamma$, see Fig.~\ref{fig:summary}.  For example, though particle irradiation has successfully doubled the critical current in cuprates and certain iron-based superconductors, the same ions and energies do not even produce universal effects in materials belonging to the same class of superconductors\cite{Tamegai2012}.

Though we can indeed tune the disorder landscape, we certainly do not have full control of it.  Defects such as stacking faults, twin boundaries, and dislocations are often intrinsic to materials and their densities are challenging to tune. As a further complication to understanding vortex-defect interactions, superconductors often have mixed pinning landscapes, i.e., containing multiple types of defects. Though these landscapes immobilize vortices over a broader range of conditions (temperatures and fields) than landscapes containing only one type of defect, it is challenging to infer the vortex structures that form within these materials and no techniques currently exist to fully image these structures and vortex-defect interactions on a microscopic level.

Generally speaking, achieving a materials-by-design approach first entails garnering a sufficient microscopic understanding of vortex-defect and vortex-vortex interactions, then incorporating these details into simulations. Significant headway has been made along these lines with the implementation of large-scale time-dependent Ginzburg-Landau (TDGL) simulations to study vortex motion through disordered media. Spearheaded by the Argonne National Laboratory, this effort has accurately modeled critical currents $\Jc$ in thin films (2D), layered and anisotropic 3D materials, as well as isotropic superconductors~\cite{Sadovskyy2015a,Sadovskyy2016a,Glatz2016,Sadovskyy2017,kimmel2019}. 
Additionally, it has determined the optimal shape, size, and dimensionality of defects necessary to maximize $\Jc$, depending on the magnitude and orientation of the magnetic field~\cite{Koshelev2016,Sadovskyy2016b,Kimmel2017,Sadovskyy2019}. Backed by good agreement with experimental and analytic results for simple geometries \cite{Willa2015a, Willa2015b, Willa2016, Willa2018c}, the utility of the numerical routine has successfully been extended to previously unknown territories, optimizing pinning geometries outside the scope of analytic methods \cite{Glatz2016, Kimmel2017, Koshelev2016, Kwok2016,Papari2016, Sadovskyy2015a, Sadovskyy2016a, Sadovskyy2016b, Sadovskyy2019, Willa2018b,ted100}. In fact, these TDGL simulations have unveiled new phenomena---such as a small peak in $\Jc(B)$ at high fields that is caused by double vortex occupancy of individual pinning sites.\cite{Willa2018a}  The Argonne team has even deployed mature optimization processes based on targeted evolution using genetic algorithms.~\cite{Sadovskyy2019}  This is a remarkable step towards the goal of \textit{critical-current-by-design}.

A critical-current-by-design must consider thermal fluctuations, which dramatically impact the critical current due to the effects of rapid thermally-induced vortex motion (thermal creep). Creep, which manifests as a decay in the persistent current over time, is rarely problematic in low-$\Tc$ superconductors as it is typically quite slow. Consequently, Nb--Ti solenoids in magnetic resonance imaging systems can operate in \textit{persistent mode}, retaining a fairly constant magnetic field for essentially indefinite time periods. However, creep is fast in high-$\Tc$ superconductors, restricting applications and reducing the effective $\Jc$.

For the sake of power and magnet applications, the goals are clear---maximize the critical current and minimize creep. Regarding the former, there is much room for improvement: no superconductor containing vortices has ever achieved a $\Jc$ higher than 25\% of its theoretical maximum, which is thought to be the depairing current $\Jdp=\Phi_0/(3\sqrt{3}\pi\mu_0 \xi \lambda^2$).  Regarding creep, we are fighting a theoretical lower bound.\cite{Eley2017a} This lower bound positively correlates with a material's Ginzburg number $Gi = (\gamma^2/2)(\kB \Tc/ \varepsilon_{sc})^2$, which is the ratio of the thermal energy to the superconducting condensation energy $\varepsilon_{sc} = (\Phi_{0}^{2} / 2 \pi \mu_{0} \xi^{2} \lambda^{2}) \xi^{3}$. The implications are grim: creep is expected to be so fast in potential, yet-to-be-discovered room-temperature superconductors rendering them unsuitable for applications. The caveat is that this lower bound is limited to low temperatures and fields (single vortex dynamics), and collective vortex dynamics could be key to achieving slow creep rates.

Though superconducting sensing and computing applications do not require high currents, vortices still pose a nuisance by limiting the lifetime of the quantum state in qubits \cite{Oliver2013}, inducing microwave energy loss in resonators \cite{Song2009a}, and generally introducing noise.  It is known that dissipation from vortex motion reduces the quality factor in superconducting microwave resonators, which are integral components in certain platforms for quantum sensors and the leading solid-state architecture for quantum computing (circuit-QED)\cite{Wallraff2004, Blais2004, Krantz2019, Muller2019}.  They are used to address and readout qubits as well as mediate coupling between multiple qubits. Consequently, resonator stability can be essential for qubit circuit stability.  Moreover, thermally activated vortex motion can contribute to $1/f$ noise and critical current fluctuations \cite{Trabaldo2019, VanHarlingen2004} in quantum circuits and is a suspected source of the dark count rate in superconducting nanowire single-photon detectors \cite{PhysRevB.83.144526, Yamashita2013}.

In these quantum circuits, vortices appear due to pulsed control fields, ambient magnetic fields\cite{Song2009}, and the self-field generated by bias currents \cite{Yamashita2013}.  Mitigating the effects of vortices requires heavy shielding to block external fields and careful circuit design to control their motion, the latter of which is quite tricky.  The circuit should include structures to trap vortices away from operational currents and readout as well as narrow conductor linewidths\cite{Stan2004} to make vortex formation less favorable.  However, these etched structures may exacerbate another major source of decoherence---parasitic two-level fluctuators---defects in which ions tunnel between two almost energetically equivalent sites, which act as dipoles and thus interact with oscillating electric fields during device operation.\cite{Muller2019}  Hence, designing quantum circuits that are robust to environment noise is not trivial and has become a topic of intense interest.\cite{Muller2019, Oliver2013}

Despite all of the aforementioned application-limiting problems caused by vortices, they are not pervasively detrimental to device performance. For example, vortices can trap quasiparticles---unpaired electrons that are a third source of decoherence in superconducting quantum circuits---boosting the quality factor of resonators\cite{PhysRevLett.113.117002} and the relaxation time of qubits\cite{Wang2014}.  Furthermore, vortices can host elusive, exotic modes that are in fact useful for topological qubits, which are predicted to be robust to environmental noise that plagues other quantum device architectures. To exploit these modes in computing, we must control the dynamics of their vortex hosts.  Hence, in general, these disparate goals of eliminating or utilizing vortices for applications both require an improved understanding of vortex formation, dynamics, and, ultimately, control.

The goal of this Perspective is to present opportunities for transformative advances in vortex physics. In particular, we start by addressing vortex creep in Sec.~\ref{ssec:vortexcreep},  which notes limited knowledge of non-thermal creep processes and how recent increases in computational power will enable full consideration of creep in simulations. Second, in Sec.~\ref{ssec:Jd} we explore the true maximum achievable critical current, and the need to simultaneously exploit multiple pinning mechanisms to surpass current records for $\Jc$.  Next, Sec.~\ref{sec:RFcavities} discusses vortex-induced losses in response to AC magnetic fields and currents, with a focus on the impact on superconducting RF cavities used in accelerators and quantum circuits. We examine how the quantum revolution has handled the vortex problem for computing, while sensing applications necessitate further studies. As solving the aforementioned problems requires advanced computational algorithms, we then proceed to discuss future uses of artificial intelligence to understand the vortex matter genome in Sec.~\ref{ssec:AI}.  Finally, in Sec.~\ref{ssec:microscopy}, we recognize that most experimental studies use magnetometry and electrical transport studies to \emph{infer} vortex-defect interactions, and discuss the frontiers of microscopy that could lead to observing these interactions as well as accurately determining defect densities.

\section{Background}
Superconductors have the remarkable ability to expel external magnetic fields up to a critical value $\Hcone$, a phenomena that is known as the Meissner Effect.  Though surpassing $\Hcone$ quenches superconductivity in some materials, the state persists up to a higher field $\Hctwo=\Phi_0/2\pi\mu\xi^2$ in type-II superconductors.  In this class of materials, $\Hcone$ can be quite small (several \SI{}{\milli\tesla}) whereas $\Hctwo$ can be extremely large (from a few tesla up to as high as \SI{120}{\tesla})\cite{NMiura2002}, such that the interposing state between the lower and upper critical fields consumes much of the phase diagram and defines the technologically relevant regime.  This \textit{mixed state} hosts a lattice of vortices, whose density scales with the magnetic field $n_v \propto B$.  We should also note that, in addition to globally applied fields, self-fields from currents propagating within a superconductor can also locally induce vortices.

Each vortex carries a single flux quantum $\Phi_0$ and the core defines a nanoscale region through which the magnetic field penetrates the material. As such, the vortex core is non-superconducting, of diameter $2\xi(T)$, and surrounded by circulating supercurrents of radii up to $\lambda(T)$, as depicted in Fig.~\ref{fig:summary}. Given the dependence of the vortex size on these material-dependent parameters, vortices effectively look different in different materials---for example, they are significantly smaller in the high-temperature superconductor $\mathrm{Y}\mathrm{Ba}_2\mathrm{Cu}_3\mathrm{O}_y$ (YBCO), where $\xi(0) = \SI{1.6}{\nano\meter}$, than in Nb, in which $\xi(0)= \SI{38}{\nano\meter}$.\cite{Wimbush2015}

Vortex motion constitutes a major source of dissipation in superconductors. Propelled by currents of density $\vec{J}$ and thermal/quantum energy fluctuations, vortices experience a Lorentz force density $\vec{F}_{L} = (\vec{J} \times \vec{B})/c$ accompanied by Joule heating that weakens superconducting properties. It is this cascading process that is responsible for the undesirable impacts on applications, for which examples were provided in Sec.~\ref{sec:Introduction}.

\subsection{\label{sec:vortexpinning}Fundamentals of vortex pinning} 

\begin{figure*}[t!]
\includegraphics[width=1\linewidth]{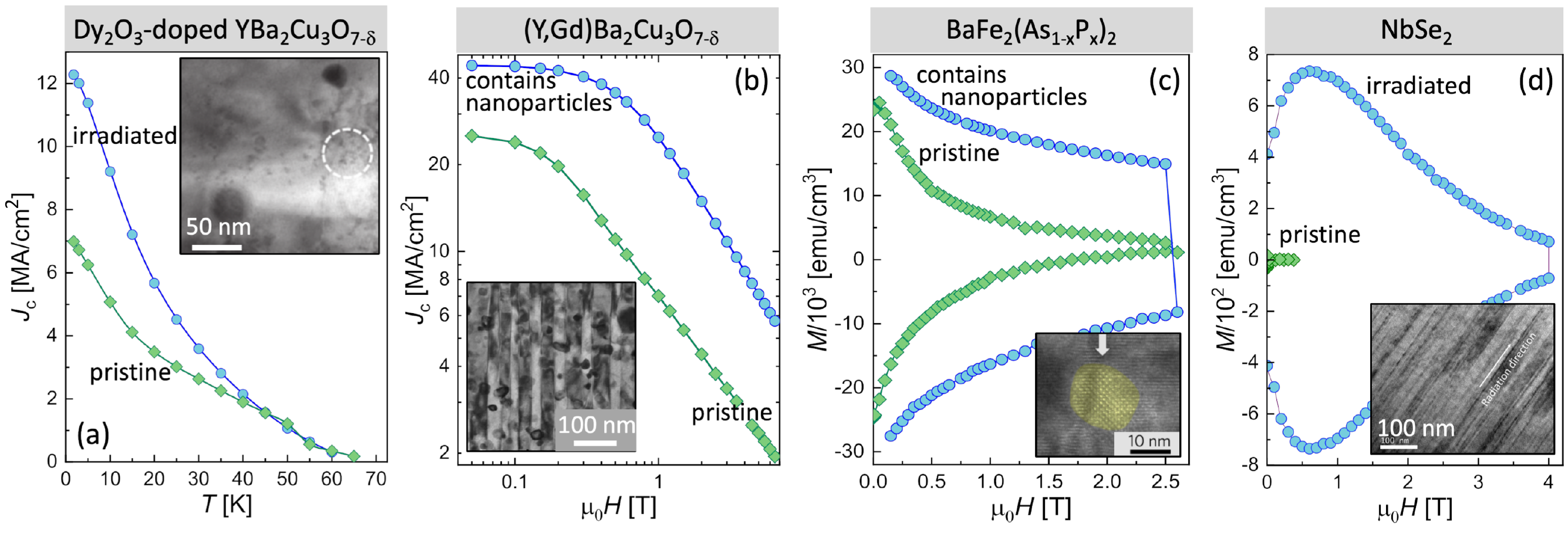}

\caption{Enhancement in $\Jc$ or $M \propto \Jc$ in (a) an oxygen-ion-irradiated Dy$_2$O$_3$-doped commercial YBCO film grown by American Superconductor Corporation\cite{Leroux2015, Eley2017a} in a field of $\SI{5}{\tesla}$, (b) a BaZrO$_3$-doped (Y$_{0.77}$Gd$_{0.23}$)Ba$_2$Cu$_3$O$_y$ film grown by Miura \textit{et al.} \cite{Miura2013k}, (c) a BaZrO$_3$-doped BaFe$_2$(As$_{1-x}$P$_x$)$_2$ film grown by Miura \textit{et al.}\cite{Eley2017}, and (d) a heavy-ion-irradiated NbSe$_2$ crystal\cite{Eley2018}. Measurements by S.\ Eley. Insets show transmission electron micrographs of defect landscape, from Refs. ~[\onlinecite{Eley2017, Miura2013k, Eley2017a, Eley2018}].}.\label{fig:Jcenhancement}
\end{figure*}

Immobilizing vortices constitutes a major research area, in which the most prominent benchmark to assess the strength of  pinning is the critical current $\Jc$. \cite{Bean1964, Zeldov1994, Zeldov1994b, Brandt1999, Willa2014, Gurevich2014, Gurevich2017, Gurevich2018, Kubo2019, Dhakal2020} Once vortices are present in the bulk, crystallographic defects such as point defects, precipitates, twin boundaries, stacking faults, and dislocations provide an energy landscape to trap vortices.  Depending on the defect type and density, one of two mechanisms are typically responsible for vortex pinning: weak collective effects from groups of small defects or strong forces exerted by larger defects.

Originally formulated by Larkin and Ovchinnikov~\cite{Larkin1979}, the theory of weak collective pinning describes how atomically small defects alone cannot apply a sufficient force on a vortex line to immobilize it. However, the collective action of many can indeed pin a vortex. In the case of a random arrangement of small, weak, and uncorrelated pinning centers, the average force on a straight flux line vanishes. Then, only fluctuations in the pinning energy (higher order correlations) are capable of producing a net pinning force.  Considering this, weak collective pinning theory phenomenology finds that the resulting critical current should scale \emph{quadratically} with the pin density $n_{p}$, i.e., $J_{c} \propto n_{p}^{2}$, see Ref.~[\onlinecite{Blatter1994}].

On the other hand, strong pinning results when larger defects each plastically deform a vortex line and a low density of these defects is sufficient to pin it.  Competition between the \emph{bare} pinning force $f_{p}$ and the vortex elasticity $\bar{C}$ generates multi-valued solutions. Because of this, a proper averaging of the effective force $\langle f_{\mathrm{pin}}\rangle$ from individual pins is non-zero and results in a critical current $J_{c} = n_{p} \langle f_{\mathrm{pin}}\rangle$. Here, the critical current reached by strong pins depends \emph{linearly} on the defect density. While conceptually simpler than weak collective pinning, it has taken significantly longer to develop a strong pinning formalism. With its completion in the early 2000s, the formalism enabled computing numerous physical observables, including the critical current~\cite{Blatter2004, Koopmann2004}, the excess-current characteristic~\cite{Thomann2012, Thomann2017,Buchacek2018, Buchacek2019a, Buchacek2019b, Buchacek2020-condmat}, and the $ac$ Campbell response~\cite{Willa2015a, Willa2015b, Willa2016, Willa2018c}.

Defects merely trap the vortex state into a metastable minimum.  Thermal and quantum fluctuations release vortices from pinning sites, and this activated motion of vortices from a pinned state to a more stable state is called vortex creep. In the presence of creep, the critical current $J_{c}$ is no longer a distinct boundary separating a dissipation-free regime from a dissipative one. Experimentally, this manifests as a power-law transition $V=J^n$ between the superconducting state in which $V=0$ and Ohmic behavior.  The creep rate $S \equiv - d \ln(J) / d\ln(t)$ then becomes $\propto 1/n$ and can be assessed by fitting the transitional regime in the current-voltage characteristic or measuring the temporal decay of an induced persistent current.  Measurements of the vortex creep rate also provide access to microscopic details such as the effective energy barriers $U^*=T/S$ surmounted and whether single vortices or bundles are creeping.

Various methods of tailoring the disorder landscape in superconductors have proven successful in remarkably enhancing the critical current. Figure \ref{fig:Jcenhancement} shows examples of cuprates, iron-based superconductors, and low-$T_c$ materials that have all benefited from incorporating inclusions.  Defects can be added post-growth, using techniques such as particle irradiation,\cite{Leroux2015, Eley2017a, Tamegai2012, Averback1997, Fang2011, Gapud2015, Goeckner2003, Haberkorn2015a, Haberkorn2015b, Haberkorn2012a, Iwasa1988, Jia2013, Kihlstrom2013, Kirk1999, Konczykowski1991, Matsui2012, Nakajima2009, Roas1989, Salovich2013, Sun2015, SwiecickiPhysRevB12, Taen2012, Taen2015, Thompson1991, Vlcek1993, Zhu1993, Leonard2013} or during the growth process by incorporating impurities to the source material.\cite{Miura2013k, Haugan2004, Horide2013, Miura2011, PalauSUST2010, WimbushSUST10} Though these processes induce markedly different disorder landscapes, both can effectuate remarkable increases in $\Jc$.  However, the conditions necessary to improve electromagnetic properties are highly material-dependent---this lack of universality renders defect landscape engineering a process of trial-and-error. 

Particle irradiation can induce point defects (vacancies, interstitial atoms, and substitutional atoms), larger random defects, or correlated disorder (e.g., amorphous tracks known as columnar defects). Notably, the critical current in commercial YBa$_2$Cu$_3$O$_{7-\delta}$ coated conductors was nearly doubled through irradiation with protons\cite{Jia2013}, oxygen-ions\cite{Leroux2015, Eley2017}, gold-ions\cite{Rupich2016}, and silver-ions.  Furthermore, iron-based superconductors have also been shown to benefit from particle irradiation\cite{Tamegai2012}.  To incorporate larger defects, such as nanoparticle inclusions, numerous groups\cite{MacManus-Driscoll2004, Feighan2017, Miura2013k, Miura2011, Miura2016, Miura2017} have introduced excess Ba and $M$ (where $M$= Zr, Nb, Sn, or Hf) into growth precursors. This results in the formation of randomly distributed 5-20 nm sized non-superconducting Ba$M$O$_3$ nanoparticles or nanorods.  This method has produced critical currents that are up to seven times higher than that in films without inclusions,\cite{Miura2013} therefore, has become one of the leading schemes for enhancing $\Jc$.

The enhancement achieved by inclusions and irradiation is often restricted to a narrow temperature and field range, partially because $\xi$ and $\lambda$ are temperature dependent, whereas the defect sizes and densities are fixed.  Another reason for the limited range of the enhancement is that, under the right conditions, certain fast moving vortex excitations may form. For example, in materials containing parallel columnar defects, double-kink excitations form at low fields and moderate temperatures that result in fast vortex creep concomitant with reduced  $\Jc$. \cite{Maiorov2009} Mixed pinning landscapes, composed of different types of defects, can indeed enhance $\Jc$ over a broader temperature and field range than inclusions of only one type and one size.  More work is indeed necessary to optimize this.

\subsection{Thermally activated vortex motion}

Vortex creep is a very complex phenomenon due to the interplay between vortex-vortex interactions, vortex-defect interactions, vortex elasticity, and anisotropy. \cite{Blatter1994, Feigelman1989}  \cite{Willa2020a}These interactions determine $U_{act}(T,H,J)$, a generally unknown regime-dependent function. The simplest creep model, proposed Anderson and Kim, neglects the microscopic details of pinning centers and considers vortices as non-interacting rigid objects hopping out of potential wells of depth $U_{act}\propto U_{p}|1 - J/J_c|$. However, as elastic objects, the length of vortices can increase over time under force from a current and vortex-vortex interactions are non-negligible at high fields. As such, the Anderson-Kim model's relevance is limited to low temperatures and fields. 

At high temperatures and fields, collective creep theories, which consider vortex elasticity, predict an inverse power law form for the current-dependent energy barrier $U_{act}(J) = U_{p}[(J_c/J)^{\mu}- 1]$, where $\mu$ is the so-called glassy exponent that is related to the size and dimensionality of the vortex bundle that hops during the creep process\cite{Blatter1991}. To capture behavior across regimes, the interpolation formula $U_{act}(J) = (U_{p}/\mu)\left[(J_c/J)^\mu - 1\right]$ is commonly used, where $\mu \rightarrow - 1$ recovers the Anderson-Kim prediction. Combining this interpolation formula with the creep time $t = t_0e^{U_{act}(J)/\kB T}$, we find the persistent current should decay over time as $J(t) = J_{c0} [1+(\mu \kB T/U_{p})\ln(t/t_0)]^{-1/\mu}$ and that the thermal vortex creep rate is
\begin{align}
    S \equiv \Big| \frac{d \ln J}{d \ln t}  \Big| = \frac{\kB T}{U_{p} + \mu \kB T \ln(t/t_0)},\label{eq:STHeqn}
\end{align}
where $\ln(t/t_0) \sim 25\text{-}30$. Because the magnetization $M \propto J$, creep can easily be measured by capturing the decay in the magnetization over time using a magnetometer. Moreover, as seen from Eq.~\eqref{eq:STHeqn}, knowledge of $S(T,H)$ provides access to both $U_{p}$ and $\mu$. Hence, creep measurements are a vital tool for revealing the size of the energy barrier, its dependence on current, field, and temperature, and whether the dynamics are glassy or plastic. It is important to note that Eq.~\eqref{eq:STHeqn} is typically used to analyze creep data piecewise---it can rarely be fit to the entire temperature range. Creep rates are not predictable and no analytic expression exists that broadly captures the temperature and field dependence of creep. Both $U_{p}$ and $\mu$ have unknown temperature dependencies, which is a major gap in our ability to predict vortex creep rates.

\subsection{Predictive vortex matter simulations}
Simulating the the behavior of vortex matter~\cite{Blatter1994,Brandt:1995,Crabtree1997,Nattermann2000,BlatterG:2003,ROPP} has a long history.  Though the value of such simulations was realized long ago~\cite{BrandtJLTP83-1,BrandtJLTP83-2}, the efficacy to produce accurate results in materials containing complex defect landscapes is considered a recent success, tied to improvements in computational power.  Specifically, we can now numerically solve more realistic models, ranging from Langevin dynamics to time-dependent Ginzburg-Landau (TDGL) equations to fully microscopic descriptions, including Usadel and Eilenberger, Bogoliubov-de Gennes, and non-equilibrium Keldysh-Eilenberger quantum transport equations. While the phenomenological TDGL equations describe vortex matter realistically on lengths scales above the superconducting coherence lengths, full microscopic equations are needed to describe, e.g., the vortex core accurately. This, however, means that the system sizes, which can be simulated down to the nanoscale using microscopic models, are quite limited, while TDGL can simulate macroscopic behavior including most dynamical features of vortex matter.

\begin{figure*}
\includegraphics[width=1\linewidth]{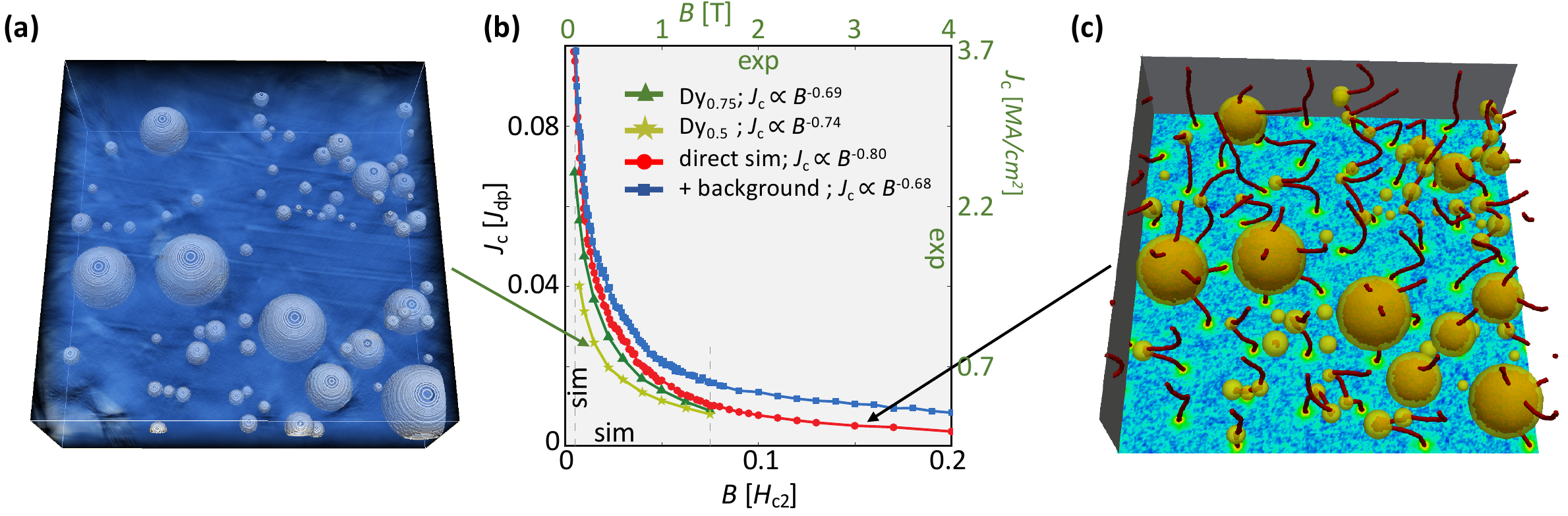}
\caption{\textbf{(a)} 3D STEM tomogram of a 0.5 Dy-doped YBCO sample. Image processing is discussed in Ref.~[\onlinecite{Leroux2015}]. \textbf{(b)} Critical current $\Jc$ as a function of the magnetic field $B$ applied along the c-axis of YBCO. The simulated field dependence (circles, red curve) with only the nanoparticles observed by STEM tomography in the sample with 0.5 Dy doping exhibits almost the same exponent $\alpha$, for $J_c \propto B^{-\alpha}$, as the experiment (triangles, green curve). Adding $2\xi$ diameter inclusions to the simulation makes the dependence less steep (squares, blue curve), which yields an exponent very similar to the experimental one in the sample with 0.75 Dy doping (stars, yellow curve).
\textbf{(c)} Snapshot of the TDGL vortex configuration with applied magnetic field and external current for the same defect structure as in the experiment (a). Isosurfaces of the order parameter close to the normal state are shown in red and follow both vortex and defect positions. The amplitude of the order parameter is represented on the backplane of the volume where blue corresponds to maximum order parameter amplitude. Arrows indicate the experimental and simulated $\Jc$ dependencies.}\label{fig:tomogram}
\end{figure*}

The Langevin approach only considers vortex degrees of freedom, while mostly neglecting elasticity and vortex-vortex interactions, which are nonlocal effects. Hence, its accuracy is limited to when inter-vortex separations are significantly larger than $\xi$, vortex pinning sites are dilute, or the superconducting host is sufficiently thin that vortices can be considered 2D particles. Nevertheless, this simple picture reveals remarkably rich, dynamical behavior -- notably realizing a dependence of $\Jc$ on the strength and density of pinning centers ~\cite{BrandtJLTP83-1,BrandtJLTP83-2}, thermal activation of vortices from pinning sites~\cite{KoshelevPhysC92}, a crossover between plastic and elastic behavior~\cite{CaoPhysRevB00,DiScalaNJP12}, and dynamic ordering of vortex lattices at large velocities~\cite{KoshelevPhysRevLett94, MoonPhysRevLett96}.  

However, vortex elasticity is indeed an influential parameter in bulk systems. It results in vortex phases that are characterized by complex vortex structures, glassy phases that do not exist in 2D systems, as well as other interesting characteristics~\cite{ErtasK:1996, OtterloPRL00, BustingorryCD:2007, LuoHu:2007, LuoHuJSNM10, Koshelev:2011, DobramyslEPJ13}. Herein lies the strength of the TDGL approach, which is a good compromise between complexity and fidelity. It describes the full behavior of the superconducting order parameter~\cite{schmid} and therefore represents a `mesoscopic bridge' between microscopic and macroscopic scales. Notably, it surpasses the Langevin approach by (i) describing all essential properties of vortex dynamics, including inter-vortex interactions with crossing and reconnection events, (ii) possessing a rigorous connection to the microscopic Bardeen-Cooper-Schrieffer theory in the vicinity of the critical temperature~\cite{Gorkov:1959}, and (iii) considering realistic pinning mechanisms. Regarding pinning, it can specifically account for pinning due to modulation of critical temperature ($\delta \Tc$-pinning) or mean-free path ($\delta \ell$-pinning), strain, magnetic impurities~\cite{DoriaEPL07}, geometric pinning through appropriate boundary conditions, and, generally, weak to strong pinning regimes---all beyond the reach of the Langevin approach. Consequently, the TDGL formulation is arguably one of the most successful physical models, describing the behavior of many different physical systems, even beyond superconductors~\cite{Aranson:2002}.

In its early days, the TDGL approach was used to study depinning, plastic, and elastic steady-state vortex motion in systems containing twin and grain boundaries as well as both regular and irregular arrays of point or columnar defects.~\cite{kaper,crabtree2000} Those simulations were predominately used to illustrate the complex dynamics of individual vortices because computational limitations prohibited the study of large-scale systems with collective vortex dynamics. Only later did simulation of about a hundred vortices in two-dimensional systems become possible, resulting in predictions for, e.g., the field dependence of $\Jc$ in thin films with columnar defects~\cite{Palonen2012}.

A 2002 article by Winiecki and Adams~\cite{Winiecki:2002} deserves credit as one of the first simulation-based studies of vortex matter in three-dimensional superconductors that produced a realistic electromagnetic response. Later, in 2015, Koshelev et al.~\cite{Koshelev2016} achieved a major technical breakthrough by investigating optimal pinning by monodispersed spherical inclusions. The simulated system size of $100\xi \! \times \! 100\xi \! \times \! 50\xi$ was much larger than any previously studied system, enabling even more realistic simulations of the collective vortex dynamics than previous works.
Their computational approach is based on an optimized parallel solver for the TDGL equation~\cite{sadovskyy+jcomp2015}, which allows for simulating vortex motion and determining the resulting electrical transport properties in application-relevant systems. The efficacy of this technique is best demonstrated in a study~\cite{Sadovskyy2016a} that applied the same approach to a `real' pinning landscape by incorporating scanning transmission electron microscopy tomography data of Dy-doped YBa$_2$Cu$_3$O$_{7-\delta}$ films~\cite{Ortalan:2009, Herrera2008}, and the results showed almost quantitative agreement of the field and angular dependent critical current with experimental transport measurements, see Fig.~\ref{fig:tomogram}.

Finally, we discuss applying TDGL calculations to commercial high-temperature superconducting tapes, which typically consist of rare earth (RE) or yttrium barium copper oxide (REBCO) matrices. Specifically, Ref.~[\onlinecite{Sadovskyy2016b}] simulated vortex dynamics in REBCO coated conductors containing self-assembled BaZrO$_3$ nanorods, and reported good quantitative match to experimental measurements of $\Jc$ versus the applied magnetic field-angle $\theta$.  
Most notably, the simulations demonstrated the non-additive effect of defects: adding irradiated columnar defects at a 45$^\circ$ angle with the nanorod (c-) axis removes the $\Jc(\theta=0^\circ)$ peak of the nanorods and generates a peak at $\theta=45^\circ$ instead. This study then went beyond simply reproducing experimental behavior, and predicted the optimal concentrations of BaZrO$_3$ nanorods that are necessary to maximize $\Jc$, which it found to be 12-14\% of $\Jdp$ (at specific $\theta$)---far higher than had been experimentally achieved in similar systems.  This approach is certainly more efficient than the standard trial-and-error approach, growing and measuring samples with a large variety of defect landscape.

These recent successes in accurately predicting $\Jc$ in superconductors based on the microstructure highlight how close we are to the ultimate goal of tailoring pinning landscapes for specific applications with well-defined critical current requirements.  Constituting the new \textit{critical-current-by-design} paradigm,~\cite{ROPP,ted100} the routine use of TDGL simulations for efficient defect landscape optimization is a transformative opportunity in vortex physics, as is expanding these computational successes to include the use of artificial intelligence algorithms.  Furthermore, microscopic and far-from-equilibrium simulations of vortex matter beyond the TDGL approach require significant computational resources and are only now becoming feasible. We will discuss related developments in Sec.~\ref{ssec:AI}.

\section{Transformative Opportunities}

\subsection{Vortex Creep\label{ssec:vortexcreep}}

In this section, we identify major opportunities to accelerate our understanding of thermally-activated vortex hopping (thermal creep) and non-thermal tunneling (quantum creep) between pinning sites. Only limited situations are amenable to an analytic treatment of vortex creep: these include thermal depinning of single vortices and vortex bundles in the regime of weak collective pinning.  In the strong pinning regime, e.g., for columnar defects, we must consider complicated excitations that form during the depinning process.  Activation occurs via half-loop formation\cite{PhysRevB.51.6526}, which is depicted in Fig.~\ref{fig:vortexstructures}.  During this process, the vortex nucleates outside of its pinned position, and the curved unpinned segment grows over time as a current acts on it, until the entire vortex eventually leaves the pinning site.  Because half-loop formation likely occurs in a range of high-current-carrying materials, which may contain amorphous tracks, nanorods, or twin boundaries, numerical treatment of vortex creep within the strong pinning framework is of significant interest.

The first task involves studying creep of isolated vortices, pinned by a single strong inclusion or columnar defect. In accordance with analytic predictions, an increase in temperature shifts the characteristic depinning current below $\Jc$, rounds the $IV$ curves and affects the excess-current characteristic far beyond $\Jc$\cite{Buchacek2019a, Buchacek2019b, Buchacek2020-condmat}. The next steps will involve studying multiple vortices, more defects, and mixed defect landscapes, which will indeed increase the complexity of the problem, warranting computational assistance.

Recent advances in computational power and high-performance codes will enable tackling these challenges, which involve long simulation times at exponentially low dynamics. Instead of simulating the thermal relaxation of a metastable configuration in a single 'linear' simulation, the same configuration can be simulated in parallel, i.e., experiencing fluctuations along different 'world lines'. This accelerates the search for a rare depinning event, after which parallel computations are interrupted and restarted from new depinned configurations.

\begin{figure*}
\includegraphics[width=1\linewidth]{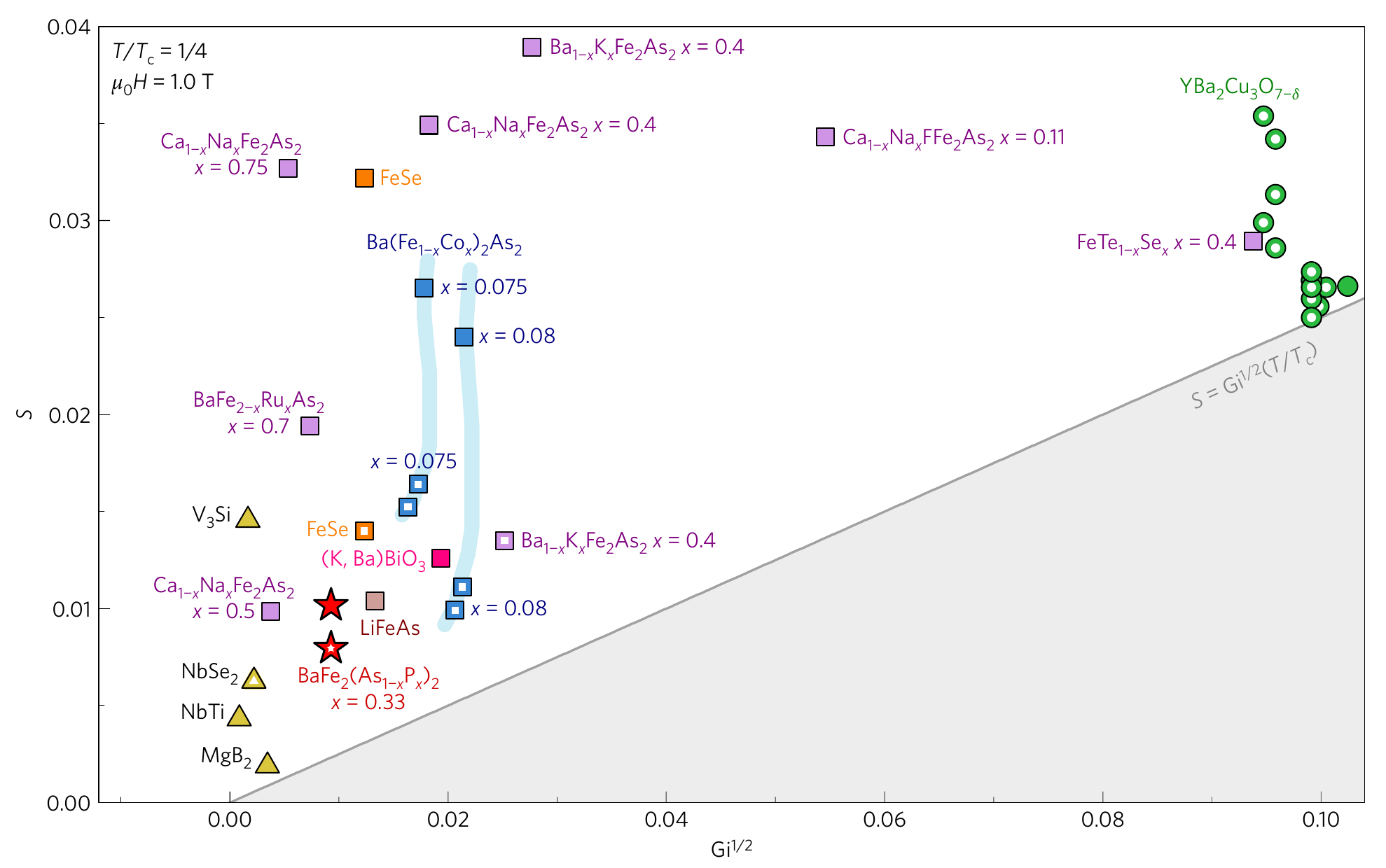}
\caption{Creep at reduced temperature $T/T_c$ = 1/4 and a field of $\mu_0H = 1 \textnormal{ T}$ for different superconductors plotted versus $Gi^{1/2}$. The open symbols indicate materials for which the microstructure has been modified either by irradiation or incorporation of inclusions. The solid grey line represents the limit set by $Gi^{1/2}(T/T_c)$.  The result predicts that the creep problem even in yet-to-be-discovered high-$T_c$ superconductors may counteract the benefits of high operating temperatures. Material from S. Eley, et al.\ Nat.\ Mater.\ 16, 409–413 (2017). Copyright 2017, \emph{Nature Publishing Group.}
}\label{fig:SvsGi}
\end{figure*}

In a 2017 paper\cite{Eley2017a}, we found that the minimum achievable thermal creep rate in a material depends on its Ginzburg number $Gi$ as $S \sim Gi^{1/2}(T/T_c)$, shown in Fig.~\ref{fig:SvsGi}.  Our result is limited to the Anderson-Kim regime and considered pinning scenarios with analytically determined pinning energies $U_P$.  It also somewhat gravely predicts that there is a limit to how much creep problem in high-$\Tc$ superconductors, which tend to have high $Gi$, can be ameliorated, such that we may expect the performance of yet-to-be discovered room temperature superconductors to be irremediably hindered by creep. However, YBCO films containing nanoparticles demonstrate non-monotonic temperature-dependent creep rates $S(T)$, such that $S$ dips to unexpectedly low values at intermediate temperatures outside of the Anderson-Kim regime\cite{Eley2017a}.  This dip, thought to be induced by strong pinning from nanoparticles, suggests that collective pinning regimes may hold the key to inducing slower creep rates that dip below our proposed lower limit in the Anderson-Kim regime.  A numerical tacking of the vortex creep problem would improve our theoretical understanding of creep and answer a major open question in vortex physics -- what is indeed the slowest achievable creep rate in different superconductors?

Our finding of the lower limit to the creep rate reduces the guesswork in trial-and-error approaches to optimizing the disorder landscape and improves our ability to select a material for applications requiring slow creep. Yet, ultimately, a material's quantum creep rate actually sets its minimum achievable creep rate.  This is a regime that is has received relatively little attention---there have been few theoretical and experimental studies of quantum creep. Theoretical models are limited to tunneling barriers induced by weak collective pinning\cite{Blatter1991, PhysRevB.47.2725} and columnar defects,\cite{PhysRevB.51.1181} though most materials have very complex, mixed pinning landscapes. Most experimental work has focused on cuprates, determining a crossover temperature of $\sim$ 8.5-11 K in YBCO films,\cite{PhysRevB.64.094509, LANDAU2000251, Luo_2002}, 1.5-2 K in YBCO crystals,\cite{PhysRevB.59.7222, LANDAU2000251} 5-6 K in Tl$_2$Ba$_2$CaCu$_2$O$_8$ films,\cite{PhysRevB.59.7222, PhysRevB.47.11552}, 17 K in TlBa$_2$CaCu$_2$O$_{7-\delta}$,\cite{PhysRevB.64.094509} 30 K in HgBa$_2$CaCu$2$O$_{6+\delta}$.\cite{PhysRevB.64.094509} Klein et al.\cite{PhysRevB.89.014514} studied an iron-based superconductor, finding crossover around 1 K in Fe(Te,Se).   No studies have been conducted in materials containing inclusions nor using any systematic tuning of the energy barrier.

Furthermore, the crossover between the thermal and quantum creep is unclear. As previously mentioned, the Anderson-Kim model's relevancy is limited to low temperatures $\kB T \ll U_{p}$ in which $S$ is expected to increase approximately linearly with temperature. A linear fit to this regime often extrapolates to non-zero $S$ at $T = 0$, suggestive of non-thermal creep. In fact, it is common to perfunctorily attribute this extrapolation to quantum creep without conducting measurements in the quantum creep regime. However, there are compelling discrepancies between typical experimental results in this context and theory. For example, theory predicts that the tunneling probability should decrease with bundle size, whereas experiments often observe the opposite trend (positive correlation between low temperature $S$ and field)\cite{Lykov2013}. Theory also predicts a quadratic, rather than linear, temperature-dependent $S(T \rightarrow 0)$\cite{Lykov2013, PhysRevB.59.7222}. That is, quantum creep may be thermally assisted\cite{Blatter1991}, and not simply present itself as a temperature-independent creep rate at low temperatures. An even more confounding result is that Nicodemi et al.\cite{PhysRevLett.86.4378} predicted non-zero creep rates at $T = 0$ using Monte Carlo simulations based on a purely classical vortex model and reconciled it with non-equilibrium dynamics.

It has also been suggested that the overall measured creep rate is simply the sum of the thermal and quantum components.\cite{PhysRevB.64.094509}  However, in some iron-based superconductors,\cite{Haberkorn2012a, Haberkorn2014} $S$ is fairy insensitive to $T$ or even decreases with increasing $T$ up to fairly high fractions of $\Tc$. Hence, either quantum creep is a significant component at surprisingly high temperatures or the creep rate dramatically decreases at temperatures below the measurement base temperature, motivating the need for lower temperature creep measurements.   Superconductors with high normal-state resistivity $\rho_n$ and low $\xi$, such as high-$\Tc$ cuprates, are the best candidates for having measurable quantum creep rates.  This is because the effective quantum creep rate is predicted to be
\begin{align}
\!\!\!
S_q = 
\begin{cases}
   -(e^2 \rho_n / \hbar \xi) \sqrt{\Jc / \Jdp},& \!\!\!\!\text{if } L_c<a_0 \\
   -(e^2 \rho_n / \hbar \lambda)(a_0/\lambda)^4(a_0/\xi)^9(\Jc/\Jdp)^{9/2}, & \!\!\!\!\text{if } L_c > a_0
\end{cases}\!\!
\end{align}
where $L_c$ is the length of the vortex segment (or bundle) that tunnels~\cite{Blatter1994}.  Determining the dependence of the quantum creep rate on material parameters in superconductors would fill a major gap in our understanding of vortex physics. This would significantly contribute towards a comprehensive model of vortex dynamics, and reveal whether creep may induce measurable effects in quantum circuits, which typically operate at millikelvin temperatures.

\subsection{Pinning at the extreme: Can the critical current reach the depairing current?\label{ssec:Jd}}

Cooper pairs constituting the dissipationless current in superconductors will dissociate when their kinetic energy surpasses their binding energy. Theoretically, this could be achieved by a sufficiently high current, termed the \textit{depairing current}, $\Jdp$. Consequently, $\Jdp$ is recognized as the theoretical maximum achievable $\Jc$, such that $\Jc/\Jdp$ is often equated with the efficiency $\eta$ of the vortex pinning landscape, which may be confusing as a perfect defect would not produce $\Jc=\Jdp$.\cite{Wimbush2015}

\begin{figure}
\centering
\includegraphics[width=1\linewidth]{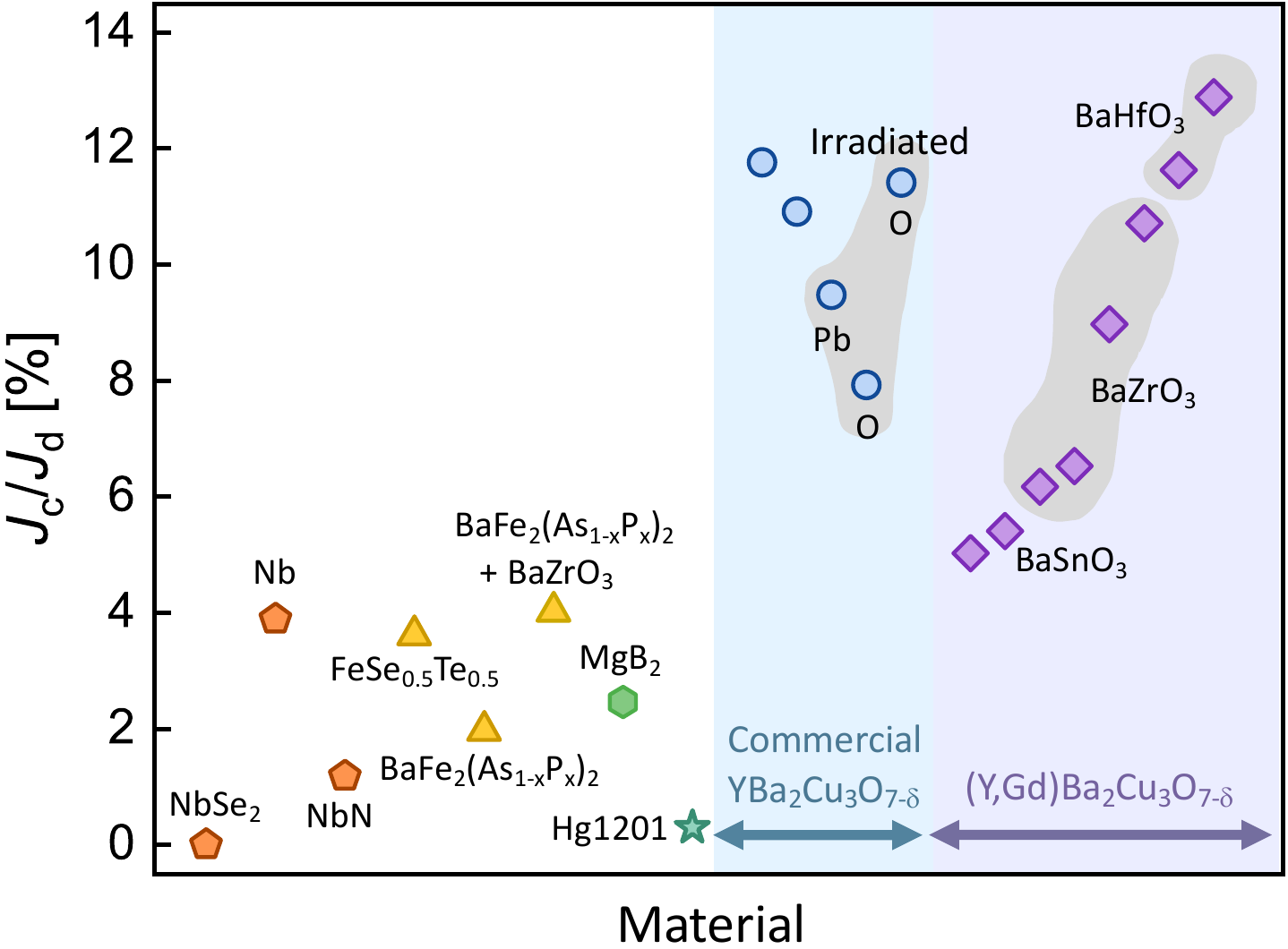}
\caption{\label{fig:JcJd} Critical current $\Jc$ normalized to the depairing current for various superconductors at $T=\SI{5}{\kelvin}$ and $\mu_0 H = \SI{0.3}{\tesla}$.  The data includes Dy$_2$O$_3$-doped YBa$_2$Cu$_3$O$_{7-\delta}$ commercial coated conductors and B$M$O$_3$-doped Y$_{0.77}$Gd$_{0.33}$Ba$_2$Cu$_3$O$_{7-\delta}$ films (where $M =$ Sn, Zr, or Hf), all grown via metal organic deposition.}
\end{figure}

The most successful efforts to carefully tune the defect landscape obtain $\Jc/\Jdp$ of only 20-30\%.\cite{Civale10201, Selvamanickam_2015} As exemplified by a series of samples we have measured, see Fig.~\ref{fig:JcJd}, most samples produce $\Jc / \Jdp < 5\%$, whereas $\Jc / \Jdp$ is routinely higher for coated conductors (REBCO films).  Though this at first appears to be a far cry from the ultimate goal, some surmise that this is indeed near the maximum that can be achieved by immobilizing vortices by means of \textit{core pinning}, which merely refers to a vortex preferentially sitting in potential wells defined by a defect to minimize the energy of its core.  Wimbush et al.\ \cite{Wimbush2015} present a compelling argument that core pinning can obtain a maximum $\Jc/\Jdp$ of only $30 \%$---A current equivalent to $\Jdp$ would produce a Lorentz force $f_d = \Jdp \Phi_0 = 4 \Hc \Phi_0 / 3 \sqrt{6}\mu_0 \lambda$, with $\Hc =\Phi_0/2\sqrt{2} \pi \lambda \xi$ the thermodynamic critical field. At the same time, the condensation energy $\varepsilon_{sc}$ produces a characteristic pinning force $f_p^{core} \sim \varepsilon_{sc}/\xi \approx \pi \xi^2 \Hc^2 / 2\mu_0$, such that the ratio of the maximal core pinning force to the depairing Lorentz force is
\begin{align}
    f_p^{core} / f_d &= 3 \sqrt{3} / 16 \approx 32 \%.
\end{align}
Similarly, Matsushita \cite{matsushita2007} performed a more precise calculation by considering the effects of the geometry of the flux line, and found that $f_p^{core}/f_d \approx 28 \%$.  Hence, decades of work in designing the defect landscape to pin vortex cores may have indeed nearly accomplished the maximum efficiency achievable by means of core pinning.

If the ultimate goal of $\Jc = \Jdp$ cannot be obtained by core pinning alone, are there other mechanisms to immobilize vortices that could indeed produce $\Jc/\Jdp > 30 \%$? Magnetic interactions between vortices themselves or vortices and magnetic inclusions can also restrict the motion of a vortex---referred to as \textit{magnetic pinning}. Herein lies a transformative opportunity to make large strides towards approaching $\Jc = \Jdp$. Magnetic pinning alone or in combination with core pinning may produce unprecedentedly high values for $\Jc$, though this mechanism has received considerably less attention than core pinning because it is quite complicated to actualize.

A high density of vortices tend to arrange themselves into a hexagonal lattice, and one pinned to a defect via core pinning may restrict the motion of its neighbors, subsequently affecting its neighbors' neighbors due to magnetic vortex-vortex interactions, which occur over a length scale of $\lambda$.  A magnetic inclusion provides another opportunity for inflicting magnetic and core pinning on a vortex. Again, following the arguments of Wimbush\cite{Wimbush2015}, we can compare the pinning force induced by core pinning to that of magnetic pinning. The magnetic Zeeman energy $\varepsilon_{mag} = \frac{1}{2} \int_{A} M \cdot B \,dA$ produced by a strong ferromagnet is much greater than the condensation energy and may be several orders of magnitude greater than the core pinning energy. However, it is unclear whether the resulting pinning force is greater because it occurs over the longer length scale of $\lambda$ versus $\xi$, i.e., $f_p^{mag} \sim \varepsilon_{mag} / \lambda$, such that
\begin{equation}
   f_p^{mag} / f_p^{core} \approx 2 \sqrt{2} (\mu_0 M / \Hc). 
\end{equation}
Hence, the advantage depends on the ratio of the magnetization of the pinning site to the thermodynamic critical field.  Independent of whether $f_p^{mag}$ surpasses $f_p^{core}$, concomitant mechanisms would produce an additive effect that may surpass current record values of $\Jc$. Yet, ferromagnets in proximity to superconductors can locally degrade superconductivity by inducing pair breaking, such that it is challenging to incorporate ferromagnetic vortex pinning centers without compromising the superconducting state. This complication combined with the typical materials science considerations of incorporating inclusions that do not induce too much strain on the surrounding superconducting matrix will make this a challenge all but insurmountable.

\begin{figure}
\centering
\includegraphics[width = 0.45\textwidth]{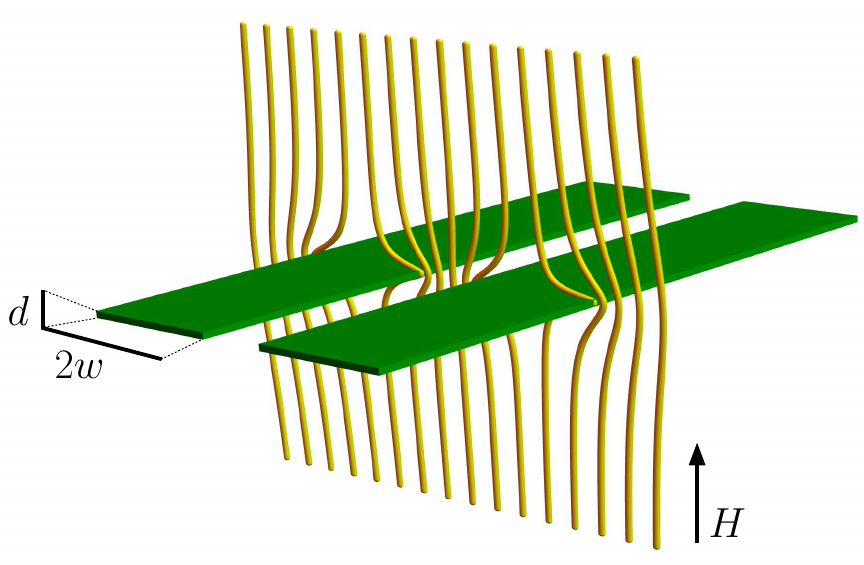}
\caption{\label{fig:doublestrip} Field profile in a double-strip geometry before penetration of vortices \cite{Willa2014}. The arrangement and geometry (e.g.\ width $w$ and thickness $d$) of the specimen significantly influence the relative importance of Bean-Livingston, geometric and bulk pinning. Material from R.\ Willa, \emph{ETH Zurich Research Collection}, see Ref.~[\onlinecite{Willa2016-thesis}].
}
\end{figure}

In addition to magnetic pinning, exploiting geometric pinning provides another potentially transformative opportunity to dramatically boost $\Jc$ in superconductors. In clean, narrow (sub-micron) superconducting strips, geometric restrictions can induce self-arrest of vortices recovering the dissipation-free state at high fields and temperatures due to surface/edge (Bean-Livingston barrier) \cite{Bean1964} or geometric \cite{Zeldov1994, Zeldov1994b, Brandt1999, Willa2014} pinning. Figure~\ref{fig:doublestrip} depicts an example of geometric vortex pinning around two superconducting strips.

Moreover, at a fixed applied current, the magnetoresistance (MR) shows oscillations with increasing magnetic field, indicating the penetration of complete vortex rows into the system~\cite{Papari2016}. Therefore, these MR oscillations are a way to determine the vortex structure in nanoscale superconductors. At very high fields, the vortex lattice in these strips starts to melt. Combining magnetoresistance measurements and numerical simulations can then relate those MR oscillations to the penetration of vortex rows with intermediate geometrical pinning, where the vortex lattice remains unchanged, and uncover the details of geometrical melting. This opens the possibility to control vortices in geometrically restricted nanodevices and represents a novel technique of `geometrical spectroscopy'. Combined use of MR measurements and large-scale simulations would reveal detailed information of the structure of the vortex system. A similar re-entrant behavior was observed in superconducting strips in a parallel field configuration: Here, high fields lead to `vortex crowding', in which a higher density of vortex lines starts to straighten, therefore reducing the Lorenz force on the vortices. The result is an intermediate dissipation-less state~\cite{parallel2017}.

The situation becomes more complex when one considers nanosized superconducting strips and bridges, in which vortex pinning is dictated by an intricate interplay of surface and bulk pinning. As described above, in the case of a very narrow bridge, $\Jc$ is mostly defined by its surface barrier, whereas in the opposite case of very wide strips, it is dominated by its bulk pinning properties. However, understanding the intermediate regime, where the critical current is determined both by bulk pinning and by the Bean-Livingston barrier at the edge of a strip is of great interest in small superconducting structures and wires. 

Recent studies~\cite{kimmel2019} revealed that while bulk defects arrest vortex motion away from the edges, defects in their close vicinity promote vortex penetration, thus suppressing the critical current. This phenomenon is also quite important in the study of superconducting radio-frequency cavities. Furthermore, the role of defects near the penetrating edge is asymmetric compared to the exit edge of a superconducting strip. This complex interplay of bulk and edge pinning open new opportunities for tailored pinning structure for a given application.
In the simple case of the straight strip with similar-type spherical defects, an optimized defect distribution can have a more than 30\% higher critical current density than a homogeneously disorder superconducting film.

The need for high-current, low-cost superconductors continues to grow with new applications in lightweight motors and generators, as well as strong magnets for high-energy accelerators, NMR machines, or even Tokamak fusion reactors. Many of these applications require large magnetic fields and therefore large critical currents, which is both a fundamental research and engineering challenge as it requires reliably fabrication of uniform, km-long high-performance superconducting cables having an optimal pinning microstructure.

Consequently, there are two main aspects that must be addressed for large-field applications:
(i) determining the best possible pinning landscape and geometry for a targeted application and
(ii) controlling fabrication of long superconducting cables to incorporate an optimized pinning landscape with highest possible uniformity. Both of these aspects are part of the critical-current-by-design paradigm~\cite{ROPP}. We will describe these in a more general context in section~\ref{ssec:AI}. Taking advantage of the modern computational approaches described there in combination with experiments opens novel pathways to new materials for large-field applications, in particular the use of high-$\Tc$ superconducting material instead of more traditional choice of elemental Nb or Nb-based compounds.

\subsection{Superconducting RF cavities and Quantum Circuits}
\label{sec:RFcavities}

\begin{figure*}
\centering
\includegraphics[width=0.8\textwidth]{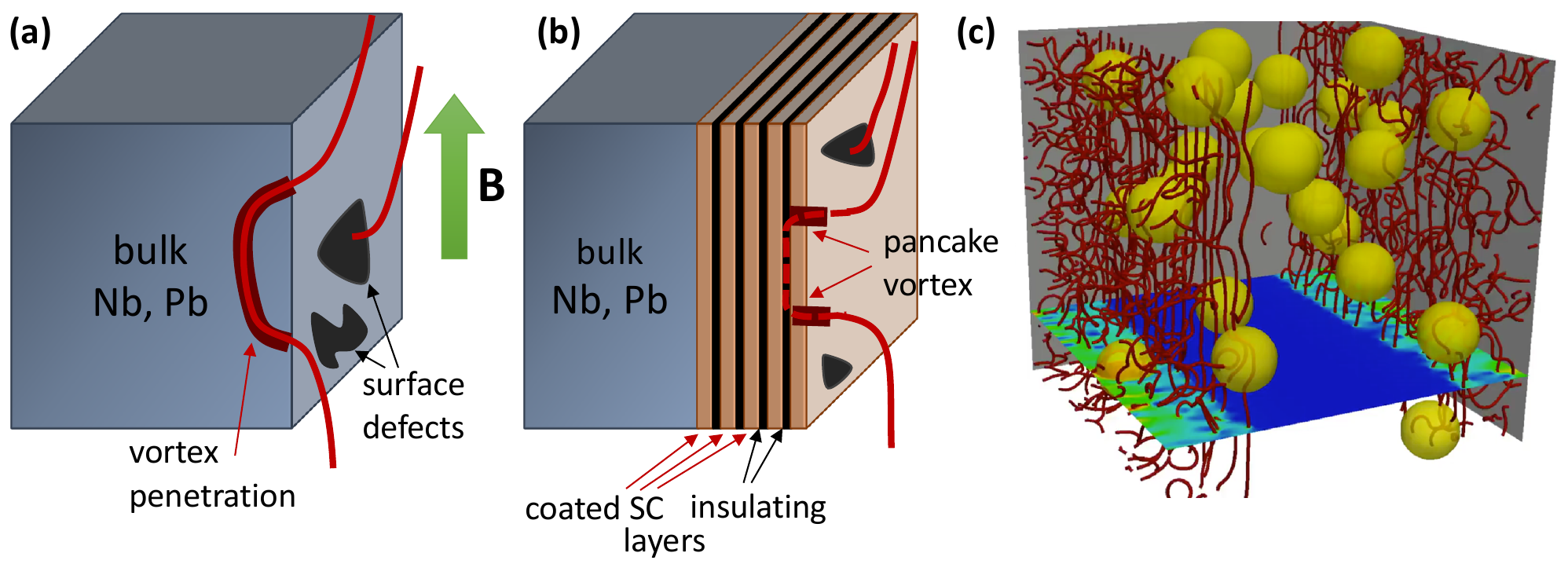}
\caption{\label{fig:srf} Surface disorder and multilayers in SRF cavities. \textbf{(a)} Sketch showing how vortices (red) parallel to the surface of a cavity wall penetrate the wall (outside the superconductor, the red lines illustrate field lines),  \textbf{(b)} intercalating insulating layers (SI[SI]S) will cause vortex pancakes to form and might limit the penetration depth of vortices~\cite{Gurevich2006,Gurevich2017}. \textbf{(c)} Simulation snapshot of surface vortex penetration into a type-II superconductor having spherical defects (yellow) near the surface in an AC magnetic field parallel to the surface. Vortex lines are shown in red. The planar projection shows the superconducting order parameter amplitude.}
\end{figure*}

\subsubsection{Superconducting RF cavities}

Most studies of vortex dynamics in superconductors are conducted using DC currents and static magnetic fields. Yet, the need to understand vortex dynamics under AC magnetic fields or AC currents is rapidly increasing, as these are the operating conditions e.g. for superconducting radio frequency (SRF) cavities and quantum circuits for sensing and computing.  Superconductors are desirable for RF devices because the minimal resistance enables very high quality factors $Q$, a metric that indicates high energy storage with low losses and narrow bandwidths.  SRF cavities are used in current and next-generation designs for particle accelerators used in, e.g., high energy physics.  In addition to $Q$, the maximum accelerating field, $E_a$, is another important metric for SRF cavity performance. The goal is often to maximize $Q$ at low drive powers, which is essential for large accelerating fields and reduced demands on cryogenic systems that are responsible for cooling the cavities. Similarly, higher $E_a$ is desirable, as it indicates larger reachable particle energies.

Elemental niobium (Nb) and Nb-based compounds are the material of choice for all current accelerator applications.  Advances in the fabrication of Nb-cavities have pushed their performance to extraordinary levels~\cite{rfbook,SRF2017}, with $Q$-values approaching $2\times 10^{11}$ and $E_a$ in excess of $45$ MV/m in Nb~\cite{qf2007}, and $E_a\sim$ 17 MV/m for Nb$_3$Sn resonators. In both cases, the magnetic field reached is above the lower critical field of the material, but below the theoretically predicted superheating field, at which vortices would spontaneously penetrate even a perfect cavity wall, shown in  Fig.~\ref{fig:srf}a.

Further increases in $E_a$ require a conceptual breakthrough in our understanding of Nb-cavity performance limits or new constituent materials. New material candidates being considered include Nb$_3$Sn, NbTiN, MgB$_2$, Fe-based superconductors, and engineered multilayer or stratified structures (see Fig.~\ref{fig:srf}b). SRF cavities operate at temperatures well below $\Tc$ and high enough frequencies to drive the superconductor into a metastable state, near breakdown. The resulting period approaches intrinsic time scales, such as vortex nucleation and quasi-particle relaxation times.

While the experimental progress in improving the performance and quality factor of SRF cavities has been impressive~\cite{SRF2017}, e.g., the counter-intuitive increase of the quality factor with nitrogen doping, it is mostly based on trail-and-error approaches. A deep fundamental understanding is important to make more systematic progress, requiring new theoretical and computational studies. Because the cavities operate out-of-equilibrium, a phenomenological description based on TDGL theory can only serve as a rough, qualitative guide. Developing a fundamental theory describing the nonlinear and non-equilibrium current response of SRF cavities requires a microscopic description based on quantum transport equations for non-equilibrium superconductors. A microscopic description, however, is challenging because the RF currents under strong drive conditions (i.e., high field frequencies and amplitudes near the breakdown/superheating field) affect both the superconducting order parameter and the kinetics of quasiparticles, all of which have to be treated self-consistently. This endeavor requires development of numerical approaches to solve the quantum transport equations, based on the Keldysh and Eilenberger formulations of non-equilibrium superconductivity in the strong-coupling limit. The Keldysh-Eilenberger quantum transport equations are, in general, non-local in space-time, non-linear, and in many physical situations involve multiple length and time scales. Solving these equations requires considerable computational resources, which are now becoming available with exa-scale computing facilities.

Herein lies a transformative opportunity to dramatically boost the performance of SRF cavities. Namely, researchers are now equipped to develop microscopic theoretical models, and incorporate them into computational codes, to reveal the origin and mechanisms that limit the accelerating field of SRF cavities.  The acquired knowledge will then guide materials optimization to maximize the critical currents, superheating fields, and quench fields. Reaching the theoretical limits for these parameters necessitates suppressing vortex nucleation.  

At high RF magnetic field amplitudes, screening currents near the vacuum-superconductor interface can nucleate Abrikosov vortices that can quench the cavity, see Fig.~\ref{fig:srf}c. This \emph{vortex breakdown} depends on (i) the amplitude and frequency of the surface field, (ii) the cavity's surface roughness, and (iii) the type, distribution, and size of defects near the interface. The impact of near-surface defects on vortices is diametric: they may reduce the potential barrier for vortex nucleation~\cite{kimmel2019}, but also pin nascent vortices generated by nucleation at the surface, preventing a vortex avalanche and substantial dissipation.  Given this, there are various possible optimal microstructures for the SRF constituent materials: (i) a ``clean'' superconductor with a maximum surface barrier, (ii) a superconductor with a thin (few $\xi$) defect-free surface and nanoscale defects in its bulk, or (iii) some special spatial gradient in the size and/or density of defects.

Large-scale TDGL simulation can be applied to study the conditions under which vortex avalanches form, devise mechanisms that are effective at mitigating these avalanches and, more generally, gain insight into the flux penetration under RF field conditions.  Furthermore, by coupling the TDGL and heat transport equations, this method can study \emph{hot spots}, providing insight on avoiding the formation of these hot spots in SRF cavity walls. As previously mentioned, though TDGL cannot produce quantitative results, it can serve as a useful guide to experiments and also provide insight into simulations based on microscopic transport equations. Lastly, though discussed in the context of SRF cavities, these new computational methods can be applied to superconducting cables for AC power applications.

\subsubsection{Vortices in Quantum Circuits\label{ssec:quantumcircuits}}
\begin{figure*}
\centering
\includegraphics[width=0.9\linewidth]{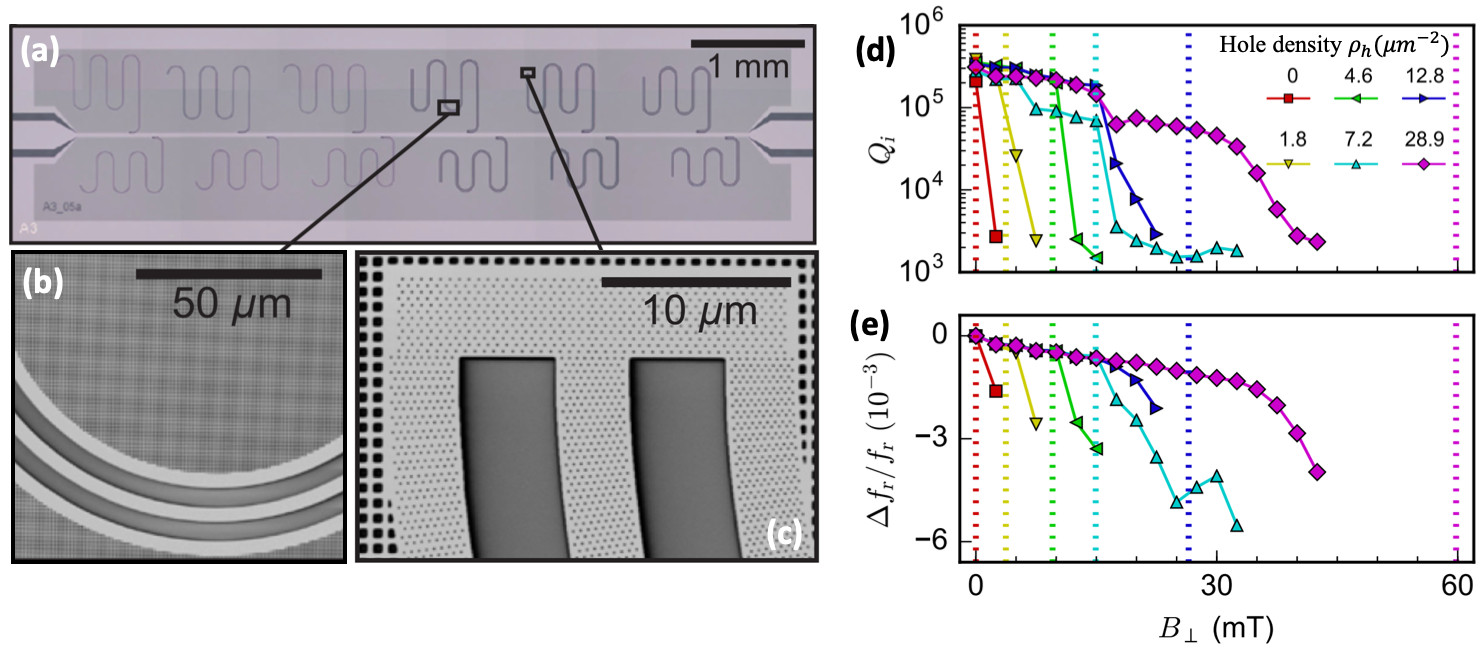}
\caption{\label{fig:Resonator}
\textbf{(a)} An optical image showing multiple $\lambda/4$ resonators multiplexed to a common feed line and surrounded by a ground plane containing holes to pin vortices and suppress vortex formation, \textbf{(b, c)} Scanning electron micrographs of superconducting CPW resonator without (\textbf{b)} and with \textbf{(c)} holes. \textbf{(d)} Quality factor $Q_i$ versus $B_{\perp}$ for varying hole density $\rho_h$. The field at which the vortex density matches the hole density (each hole is filled with one vortex) is plotted with a color-matched vertical line.  Above this threshold field, additional vortices are not pinned by the holes but instead only weakly pinned by film defects and interstitial pinning effects. \textbf{(e)} $\Delta f_r/f_r$ versus $B_{\perp}$ for varying $\rho_h$. Reprinted with permission from Ref.~[\onlinecite{Kroll2019}]. Copyright 2019, \emph{American Physical Society}.
}

\end{figure*}

\paragraph{Energy loss due to vortices.} Similar to SRF cavities, superconducting circuits for quantum information also operate at RF/microwave frequencies and are affected by vortices. Specifically, along with parasitic two-level fluctuators and quasiparticles, vortices are a considerable source of energy loss in superconducting quantum circuits\cite{Muller2019, Oliver2013, Martinis2009}. These energy loss mechanisms create a noisy environment in which the qubit irreversibly interacts stochastically, rather than deterministically. Consequently, the evolution of the quantum state is unpredictable, increasingly deviating over time from predictions until the qubit state is eventually lost.  This is called decoherence, which limits the amount of time $T_1$ over which information is retained in qubits to the microsecond range and there is typically a large spread in the $T_1$ times for each qubit in multi-qubit systems\cite{Finke2019}.

Vortices appear in superconducting quantum circuits due to stray magnetic fields, self-fields generated by bias currents, and pulsed control fields. In addition to limiting $T_1$ in qubits, thermally-activated vortex motion can cause significant noise in superconducting circuits and reduce the quality factor $Q$ of superconducting microwave resonators\cite{Song2009, Song2009a, Kroll2019}. To mitigate this, techniques have been developed to either prevent vortex formation or trap vortices in regions outside of the path of operating currents. Shielding circuits from ambient magnetic fields and narrowing linewidths constituting the device\cite{DVH2004, Samkharadze2016} significantly reduce the vortex population.  For example, for a line of width $w$, flux will not enter until the field surpasses $\Phi_0/w^2$. Because of this realization, flux qubits typically contain linewidths of $\SI{1}{\micro\meter}$, therefore exclude vortex formation up to a threshold magnetic field of roughly \SI{2}{\milli\tesla}, which is 20 times larger than the Earth's magnetic field\cite{DVH2004}.  

Though shielding has enabled remarkable headway in improving the stability of superconducting qubits for computing applications, it is not a complete solution. A reasonable amount of shielding can only suppress the field by a small amount, which may be insufficient if devices must operate in high-field environments.  Moreover, shielding may render devices useless in quantum sensing applications in which the purpose of the device is to sense the external environment. This has sparked research on further modifications to the device design and on understanding the effects of a magnetic field on different architectures of quantum circuits, including transmon qubits\cite{Schneider2019} and superconducting resonators\cite{Bothner2017, Kroll2019}, which are integral components to readout.

In addition to shielding, another common remedy for the vortex problem in superconducting circuits involves micropatterning arrays of holes in the ground plane to serve as vortex traps and reduce the prevalence of vortex formation within that perforated area\cite{Song2009a, Bothner2011, Bothner2012, Chiaro2016}, see Fig. \ref{fig:Resonator}. For example, Bothner et al.\ found that $Q(B=\SI{4.5}{\milli\tesla})$ is a factor of 2.5 higher in Nb resonators containing flux trapping holes in the ground plane \cite{Bothner2011} compared to without the holes.  However, Chiaro et al.\cite{Chiaro2016} showed that, without careful design, these features can increase the density of and subsequently losses from parasitic two-level fluctuators, thought to primarily form at surfaces and interfaces.  Moreover, coplanar waveguide resonators were recently found to be more robust to external magnetic fields when the superconducting ground plane area is reduced, which lowers the effective magnetic field inside the cavity, and by coupling the resonator inductively instead of capacitively to the microwave feedline, shielding the feedline\cite{Bothner2017}.

The methods we have discussed here engendered tremendous advances in suppressing the vortex problem in superconducting quantum circuits, however, the details are material-dependent. Likewise optimal mitigation strategies may be material-dependent. For example, Song et al.\ compared the microwave response of vortices in superconducting Re (rhenium) and Al coplanar waveguide resonators with a lattice of flux-trapping holes in the ground plane.  Generally, in both systems, vortices shift the resonance frequency $f_0$, broaden the resonance dip $|S_{21}|(f)$, and reduce the quality factor $Q$. However, vortices in the Al resonators induce greater loss and are more sensitive to flux creep effects than in the Re resonators.  The Al resonator experienced a far more substantial fractional frequency shift $df/f_0$ with increasing frequency than the Re resonator. Furthermore, while the loss $1/Q$ due to vortices increased with frequency for Re, it decreased for Al.  

Most research on the microwave response of vortices in quantum circuits is limited to Al\cite{Song2009, Song2009a, Chiaro2016, PhysRevLett.113.117002, Wang2014}, Nb\cite{Bothner2012, Bothner2011, Stan2004, Kwon2018, Golosovsky1995}, NbTiN\cite{ Samkharadze2016, Kroll2019}, and Re\cite{Song2009a}.  Whereas Al and Nb are used in commercial quantum computers, superconducting nitrides (TiN, NbN, NiTiN)\cite{Sage2011, Ohya2013, Vissers2012a, Leduc2013, Sandberg2012, Chang2013, Kerman2006, Barends2010a, Barends2010b, Bruno2015} and Re have garnered substantial attention because they may suffer less from parasitic two-level fluctuators, which are particularly problematic in oxides and at interfaces\cite{Muller2019}.  Nitrides and Re are known to develop thinner oxide layers than Al and Nb, and can be grown epitixally on common substrates\cite{Dumur2016, WangMartinis2009, Vissers2010}.  To develop a generic understanding of how to design quantum circuits that are resilient to ambient magnetic fields and control vortices in circuits made of next-generation materials, we must study circuits consisting of broader ranges of materials, perform further studies on nitride-based circuits, investigate different designs for flux trapping, and conduct imaging studies that can observe rather than infer the efficacy of vortex pinning sites. There have been a few studies that imaged vortices in superconducting strips, which provided guidance on appropriate line widths to preclude vortex formation\cite{Stan2004, Kuit2008}.  To build upon this, imaging studies (using e.g. a scanning SQUID or magnetic force microscope) of devices would inform on the efficacies of flux trapping sites, reveal locations in which vortices form, and track vortex motion.

\paragraph{Vortices in topological quantum computing schemes.}
Up until now, we have discussed vortices exclusively as a nuisance, which is indeed the case for a broad range of applications.  A notable exception lies in the burgeoning field of topological quantum computing, in which vortices serve as hosts for Majorana modes\cite{Liu2019}.  Qubits encoded using Majorana modes are predicted to be relatively robust to noise, thus have long coherence times. One way to realize this is to couple a superconductor to a topological insulator, induce vortices in the superconductor, and Majorana states are predicted to nucleate in the vortex core. (Also note that Majorana modes have been theorized to exist in other systems) \cite{Grosfeld2011, Nenoff2019, You2014, DasSarma2012, Bjorn2012, Liang2016, Alicea_2012}.  Initially elusive, signatures of Majorana vortex modes have been recently observed in a variety of systems, including the iron-based superconductor $\mathrm{Fe}\mathrm{Te}_{x}\mathrm{Se}_{1-x}$ \cite{Chiueaay2020, Ghaemi2020}, EuS islands overlying gold nanowires \cite{Manna2020}, superconducting Al layer encasing an InAs nanowire\cite{Vaitiekenaseaav2020}, and Bi$_2$Te$_3$/NbSe$_2$ heterostructures\cite{JFJia2016}. To exploit these modes for computing, we must be able to control their vortex hosts.  Consequently, vortex pinning research will be beneficial to vortex-based topological quantum computing applications.

\subsection{Vortex matter genome using artificial intelligence: Critical-current-by-design}\label{ssec:AI}

\begin{figure*}
\centering
\includegraphics[width=0.98\textwidth]{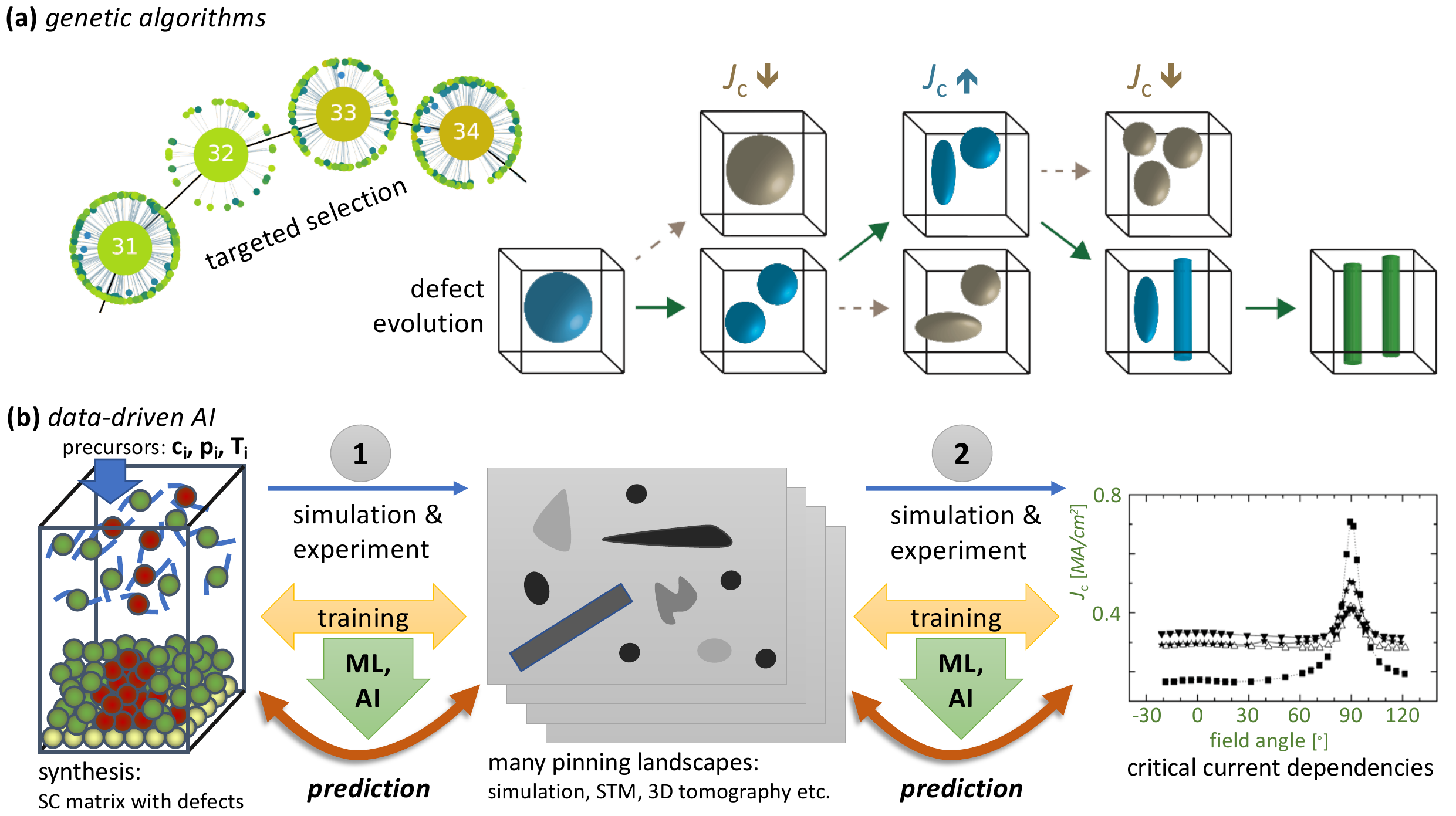}
\caption{\label{fig:AI} Critical-current-by-design using: \textbf{(a)} Genetic algorithms to optimize critical currents. Starting with a superconductor having intrinsic defects, genetic algorithms can be used to optimize the defect structure by mutation of defects and targeted selection of landscapes with larger critical currents. A mutation of a defect (or several defects at once) can be done by, e.g., translation, resizing, deletion, or splitting as sketched in the defect sequence on the right. Overall, this procedure creates ``generations'' of mutated defect configurations and only the best is selected and chosen to be the seed for the next generation shown in the partial tree on the left (circles/dots represent configurations, where the large numbered one is the best). Using neural networks and machine learning (ML) to predict the best mutations could further improve the targeted selection approach~\cite{Sadovskyy2019}. \textbf{(b)} ML/Artificial intelligence (AI) to improve and tailor defect landscapes in superconductors. \textbf{\protect\circled{1}} illustrates how AI models can be used to predict pinning landscapes from synthesis parameters and, vice versa, to predict synthesis parameters like precursor concentrations, pressures, and temperatures in, e.g., vapor deposition methods, for a targeted pinning landscape. The models need to be trained by experimental or simulation data sets. \textbf{\protect\circled{2}} similarly shows how to directly predict critical current dependencies, like field orientation dependencies, from pinning topologies and vice versa. Again, the underlying model is trained by experimental and simulation data.}
\end{figure*}

Over the years, research in superconductivity and vortex pinning has produced large amounts of experimental and simulation data on microstructures, synthesis, and critical current behavior.  More recently, artificial intelligence (AI) and machine learning (ML) approaches have enabled revolutionary advances in the fields of image and speech recognition as well as automatic translation, and are now finding an increasing number of applications in scientific research areas that deal with massive data sets, like particles physics, structural biology, astronomy, and spectroscopy.  Combining these will enable novel approaches to predict pinning landscapes in superconductors for the future design of materials with tailored properties by using sophisticated ML algorithms and AI models.  This has become a promising approach within the critical-current-by-design paradigm, which refers to designing superconductors with desired properties using sophisticated numerical methods replacing traditional trial-and-error approaches. These properties include maximizing critical currents, achieving robust critical currents with respected to variations of the pinning landscape (which is important for large-scale commercial applications), or attaining uniform critical currents with respect to the magnetic field orientation.

The next step towards advancing the use of AI/ML approaches for critical-current-by-design may be to build upon the genetic algorithms implemented in Ref.~[\onlinecite{Sadovskyy2019}] to optimize pinning landscapes for maximum $\Jc$. This approach utilizes the idea of targeted selection inspired by biological natural selection. In contrast with conventional optimization techniques, such as coordinate descent, in which one varies only a few parameters characterizing the entire sample, targeted evolution allows variations in each defect individually without any \textit{a-priori} assumptions about the defects' configuration. This essentially means that one solves an optimization problem with, theoretically, infinite degrees of freedom. Ref.~[\onlinecite{Sadovskyy2019}] demonstrated the feasibility of this approach for clean samples as well as ones with preexisting defects, e.g. as found in commercial coated conductors. The latter, therefore, provides a post-synthesis optimization step for existing state-of-the-art wires and a promising path toward the design of tailored functional materials.

However, the mutations of the defects required for the genetic algorithm [see Fig.~\ref{fig:AI}(a)] were chosen randomly. Those mutations generate ``generations'' of pinning landscapes, of which the best is chosen by targeted selection and then used as a seed for the next generation. Using a simple machine learning approach could further enhance the convergence of this method, by performing only mutations which have higher probabilities of enhancing the critical current. Besides superconductors, this methodology can be used to improve the intrinsic properties of other materials where defects or the topological structure plays an important role, such as magnets or nanostructured thermoelectrics.

Going beyond these ML-improved simulations, one can build quantitative data-driven AI approaches for superconductors that will enable, e.g. predicting the critical current phase diagram and extracting the defect morphology responsible for its performance directly from the existing accumulated experimental and simulation data without actual dynamics simulations.  Here, we will discuss two potentially transformative opportunities, summarized in Fig.~\ref{fig:AI}(b).  The first application is motivated by the need for reliably producing uniform superconductors on macroscopic commercial scales. This requires a deep understanding of material synthesis processes e.g., for self-assembled pinning structures in a superconducting matrix (see Fig.~\ref{fig:AI}(b), \textbf{\circled{1}}). Materials at the forefront of this quest are REBCO films with self-assembled oxide inclusions in the form of nanorods and nanopartices~\cite{Obradors2014, ROPP}. For example, BaZrO$_3$ (BZO) nanorods that nucleate and grow during metal organic chemical vapor deposition (MOCVD) have proven particularly effective for pinning vortices~\cite{majkic2017}.  

The major difficulties in achieving consistent and uniform critical currents in REBCO tapes containing BZO nanorods are the interplay of many parameters controlling the deposition process (temperature of the substrate and of the precursor gases, deposition rate, precursor composition, etc.) and strong sensitivity of the microstructure to small variations in these parameters.  Even for the same nominal level of Zr additives, significant variations in nanorod diameter, size distribution, spacing distribution, and angular splay have been observed. Physical factors controlling these variations remain poorly understood. For example, the nanorods’ diameter may be a mostly equilibrium property resulting from the interplay of strain and surface energies or caused by kinetic effects controlled by surface diffusion of adatoms and deposition rate. 

These complexities have precluded the development of predictive models.  However, making use of the accumulated experimental data sets and possibly synthesis/kinetic growth simulation data (Monte Carlo or molecular dynamic simulations, which are also still in an exploratory phase), allows building ML/AI models to predict pinning landscapes for given synthesis parameters as described above or, more relevant for commercial application, the prediction of synthesis parameters for a desired, uniform pinning landscape.


To constitute a complete \textit{vortex-pinning genome}, a second notable milestone is using AI to predict $\Jc$ for a given pinning landscape based solely on data recognition (disregarding TDGL simulations) and, conversely, predicting the necessary pinning landscape to produce a desired $\Jc$. In fact, the latter cannot be achieved by direct simulations.   Typical data sets, both experimental and simulation-based, contain information on defect structures, critical currents, and other transport characteristics for a wide range of magnetic fields and temperatures. Creating an organized database of this information would enable (i) quickly accessible critical current values for a wide range of conditions, (ii) an effective mapping of simulation parameters onto experimental measurements, and (iii) using the data as training sets for AI-driven predictions of defect structures for desired applications.

Experimentally, microstructures are routinely probed by transmission electron microscopy (TEM) and, less directly, by x-ray diffraction (XRD). In contrast to simulation data, which contains all information about pinning landscapes, the extracted information is usually rather limited, since TEM only allows imaging of thin slices of the material and only detects relatively large defects. A full 3D tomography of defect landscapes [cf. Fig.~\ref{fig:tomogram}(a)] is very time consuming and expensive, and therefore currently typically infeasible. The resulting AI/ML models will also allow for a cross-validation of the simulation-based data with available experimental data on materials properties in superconductors with different defect microstructures.

Overall, this AI/ML approach will directly reduce the cost and development time of commercial superconductors and, in particular, accelerate their design for targeted applications.
To estimate the benefit of such an approach, one can consider, for example, a pinning landscape defined by 9 parameters. Using traditional interpolation in this 9-dimensional parameter space one would need to have a certain number of data points per parameter. 
For the modest case of 15 data points per direction one would need to simulate (measurements are infeasible) $15^9\approx 4\cdot 10^{10}$ pinning landscapes, which -- assuming 15 minutes per simulation on a single GPU -- results in a total simulation time of a million GPU years. This simulation time is beyond current capabilities, even on the largest supercomputers. However, surrogate ML models can reduce this to approximately $10^4$ simulations, while maintaining the same resulting accuracy (see seed studies in, e.g., Ref.~[\onlinecite{crombecq2011}]).

In this section, we mentioned the complications associated with 3D tomographic imaging of a superconductor's microstructure to supply complete information for simulations. In the next section, we detail the limitations of tomographic imaging and other advanced microscopy techniques, many of which will be revolutionized by improvements in computational power and the application of advanced neural networks.  This in turn will have a transformative impact on vortex physics.

\subsection{Advanced microscopy to better understand vortex-defect interactions\label{ssec:microscopy}}
\subsubsection{Quantitative point-defect spectroscopy }

We have discussed the role of point defects in suppressing vortex motion via weak collective pinning.  Notably, the theory of weak collective pinning\cite{Larkin1979} has attracted significant attention as it can explain the origin of novel vortex phases, e.g.\ vortex glass~\cite{Fisher1989,Fisher1991} and vortex liquid~\cite{Nelson1988} phases, as well as the associated vortex melting phase transition ~\cite{Brandt1989, Houghton1989}.  It cannot, however, be used to predict $\Jc$ in single crystals, whose defect landscape is dominated by point defects.  This limitation is not necessarily reflective of gaps in weak collective pinning theory itself, but rather the fact that point defect densities are typically unknown because they are extremely challenging to measure over a broad spatial range. 

Consequently, point defects are the dark matter of materials. Herein lies yet another transformative opportunity in vortex physics.  Developing a technique to accurately measure point defect profiles and subsequent systematic studies correlating point defects, $\Jc(B,T)$, and $S(B,T)$ may lead to recipes for predictably tuning the properties of superconductors, most directly impacting crystals and epitaxial materials that lack a significant contributions from strong pinning centers. The most promising routes for quantitative point defect microscopy include scanning transmission electron microscopy (STEM), atom probe tomography (APT), atom electron tomography (AET), and positron annihilation lifetime spectroscopy (PALS). Here, we primarily focus on STEM combined with AET, then will introduce APT and PALS as other techniques with atomic-scale resolution that are relatively untapped opportunities to reveal structure-property relationships in superconductors.

In scanning transmission electron microscopy (STEM), an imaging electron beam is transmitted through a thin specimen, such that detectors can construct a real-space image of the microstructure and collect other diffraction data. In superconductors, STEM studies have revealed a panoply of defects, including columnar tracks, defect clusters, dislocations, twin boundaries, grain boundaries, and stacking faults. These studies can also provide information on other pertinent microstructural properties, including strained regions that induce variations in the superconducting order parameter, therefore, preferential regions for vortices to sit in an otherwise defect-free landscape.  To identify dopants (e.g. BaHfO$_3$ nanoparticles), STEM is also often performed in conjunction with analytical techniques, such as energy dispersive x-ray spectroscopy.

To understand the ability of STEM to determine point defect densities in superconductors, we must first understand what limits the spatial resolution and throughput. Older STEMs cannot resolve point defects due to imperfections (aberrations) in the objective lenses and other factors that set the resolution higher than the wavelength of the imaging beam.  Atomic resolution was finally achieved upon the advent of transformational aberration correction schemes, which were first successfully demonstrated in the late 1990s and have been increasingly widely adopted over the past decade\cite{Batson2002, ROSE20051, Dahmen2009, Ophus2017}.  In fact, the spatial resolution of an aberration-corrected STEM has now fallen below the Bohr radius of $\SI{0.53}{\pico\meter}$ \cite{kisielowski2008, Erni2009, Alem2009, Naoya2019}.  

Though point defects can now be imaged in superconductors, it is not straightforward to determine point defect densities. A single scan captures a small fraction of the sample, which may not be representative of defect distributions throughout the entire specimen. Accordingly, low throughput prevents collecting a sufficiently large dataset to provide a reasonably quantitative picture of defect concentrations. One of the limiting factors for throughput is the detector speed, which has recently improved significantly owing to the development of direct electron detectors such as active pixel sensors (APS) and hybrid pixel array detectors (PAD).  These detectors have higher quantum efficiency, operate at faster readout speeds, and have a broader dynamic range than conventional detectors---charge-coupled devices (CCDs) coupled with scintillators \cite{Ophus2017}.

Enabled by fast detectors, the advent of 4D-STEM\cite{ophus_2019} is another recent, major milestone that is a significant step towards determining point defect densities.  Note that 4D-STEM involves collecting a 2D raster scan in which 2D diffraction data is collected at each scanned pixel, generating a 4D dataset containing vast microstructural information. In addition to high-speed direct electron detectors, computational power was prerequisite for 4D-STEM implementation, in which massive datasets can be produced: see Ref.~[\onlinecite{ophus_2019}] for an example in which a single 4D-STEM image recorded in \SI{164}{\second} consumes \SI{420}{\giga\byte}. Hence, over the past few years, this has warranted efforts to develop fast image simulation algorithms \cite{Ophus2017} and schemes to apply deep neural networks to extract information, such as defect species and location\cite{Ziatdinov2017}.  Furthermore, STEMs can be used for electron tomography, in which images that are collected as the sample is incrementally rotated are combined to create a 3D image of the microstructure.\cite{Miaoaaf2157}

Aberration correction, high-speed detectors, and the data revolution are transformative advances that will certainly accelerate progress in understanding structure-property relationships in superconductors. Nevertheless, there are more salient impediments to an atomic-scale understanding of the true sample under study, including artifacts from sample preparation techniques\cite{SCHAFFER2012} and beam scattering within thick samples. To remedy the latter, materials are often deposited onto membranes, though this may not present a representative picture of the defect landscape when the sample is in a different form (e.g. thicker and on a different substrate).

Atom probe tomography (APT) is another microscopy technique with atomic-scale resolution, and it also provides 3D compositional imaging of surface and buried features. Over the past decade, it has become increasingly popular due to the development of a commercial local-electrode atom probe (LEAP). For APT, the specimen must be shaped as a needle and an applied electric field successively ejects layers of atoms from the surface of the specimen towards a detector.  By means of time-of-flight mass spectroscopy, the detector progressively collects information on the position and species of each atom, reconstructing a 3D tomographic image of the specimen that can span \SI{0.2 x 0.2 x 0.5}{\micro\meter} with resolution of $\SIrange[range-phrase=-, range-units=single]{0.1}{0.5}{\nano\meter}$
\cite{Petersen2011}.  As each atom is individually identifiable, it can provide remarkably revealing information on the microstructure.  
Similar to STEM, sample preparation and data processing are bottlenecks; APT also currently suffers from limited detection efficiency\cite{Kelly2007}.  Furthermore, the analyzed volume (field of view) is currently too small to be sufficiently representative of the sample to provide accurate quantitative details on point defect concentrations.  The biggest complication, however, may be that the defect landscape of the APT specimen, shaped as a needle, may dramatically differ from the material in the form in which we typically study its electromagnetic properties.

Lastly, positron annihilation spectroscopy is a hitherto untapped opportunity to correlate vacancy concentrations with electrical transport properties in superconductors. This non-destructive technique can determine information about vacancies and larger voids in a material by bombarding it with positrons at \SI{50}{\electronvolt} to \SI{30}{\kilo\electronvolt} acceleration energies,\cite{Gidley2006, Or2019} then measuring the time lapse between the implantation of positrons and emission of annihilation radiation. Upon implantation, positrons thermalize in \SI{\sim 10}{\pico\second} then either interact with an electron and annihilate or form a positronium atom (electron-positron pair)\cite{STRASKY2018455}.  Positronium atoms will then ricochet off the walls of voids and eventually annihilate, releasing a $\gamma$-ray that can be detected with integrated $\gamma$-ray detectors. The lifetime of the positron can provide information on void sizes and concentration of vacancies: longer lifetimes correspond to larger voids and higher vacancy densities.

PALS has been used for decades to sensitively detect vacancies and vacancy clusters in metals and semiconductors\cite{Schultz1988, Gidley2006} as well as probe subnanometer, intermolecular voids in polymers\cite{Pethrick1997, Gidley2006}.  Depth profiling is possible on the nm to the \SI{}{\micro\meter} scale\cite{RevModPhys.60.701, Wagner2018, Peng2005, Gidley2006} by tuning the positron implantation energy and, though some systems have beam scanning capabilities enabling lateral resolution, spatial resolution is generally quite poor due to large beam spot sizes and positron diffusion.  In most systems, the spot size is typically $\sim \SI{1}{\milli\meter}$. However, PALS instruments containing \emph{microprobes} are capable of spot sizes that are smaller than $\SI{100}{\micro\meter}$ \cite{PhysRevLett.87.067402, Gigl2017}. For example, in 2017, Gigl et al.\cite{Gigl2017} developed a state-of-the-art system with a minimum lateral resolution of \SI{33}{\micro\meter} and maximum scanning range of \SI[product-units=power]{19 x 19}{\mm}. Regarding speed, the system can scan an area of \SI[product-units=power]{1 x 1}{\mm} with a resolution of \SI{250}{\micro\meter} in less than 2 minutes, which is considered to be an exceptionally short time frame\cite{Gigl2017}. Moreover, David et al.\cite{PhysRevLett.87.067402} reported a remarkably small spot diameter of \SI{2}{\micro\meter} in a setup with a short scanning range of \SI[product-units=power]{0.2 x 0.6}{\mm}. Unfortunately, further improvements to beam focus may be ineffectual and spatial resolution comparable to electron microscopy is unreachable. The spatial resolution is ultimately limited by lateral straggle: the positron diffusion length is roughly several hundreds of nanometers in a perfect crystal, which limits the spot size even if the beam focus is improved \cite{RevModPhys.60.701}. Ongoing efforts to advance PALS include improving theoretical methods for interpretation of experimental results, advancing theoretical descriptions of positron physics (states, thermalization, and trapping), incorporating sample stages that allow tuning sample environmental conditions (e.g. temperature, biasing), and improving the efficiency of beam moderators (which convert polychromatic positron beams to monochromatic beams)\cite{RevModPhys.60.701}.

\subsubsection{Cryogenic microstructural analysis for accurate determination of structure-property relationships}

Accurately correlating the formation of different vortex structures and intricacies of vortex-defect interactions with electromagnetic response is not trivial. Typically, conventional microscopy is performed under conditions that differ from a material's actual working environment: structural characterization of superconductors is routinely conducted at room temperature whereas accessing the superconducting regime requires cryogenic temperatures and is probed using electromagnetic stimuli. Yet we know that temperature changes significantly impact the microstructure, causing strain-induced phase separation and altering defects such as dislocations.  Electromagnetic stimuli may similarly impact the defect landscape.  Hence, another transformative opportunity in vortex physics is cryogenic structural characterization of superconductors under the influence of electromagnetic stimuli, which requires advances in microscopy.

Scanning transmission electron microscopy combined with spectroscopic analysis is one of the most informative methods of gathering structural and chemical analysis at the atomic-scale.  Accurate determination of structure-property relationships require in-situ property measurements conducted concommitently with microscopy. Recent, rapid advances in in-situ transmission electron microscopy have been fueled by the introduction of a variety of commercial in-situ sample holders that allow for electrical biasing, heating, magnetic response, and mechanical deformation of nanomaterials\cite{McDowell2018}. These new capabilities have accelerated progress in a variety of fields, including battery electrochemistry, liquid-phase materials growth, bias-induced solid-state transformations in e.g. resistive switching devices for memory and neuromorphic applications, gas-phase reactions and catalysis, solid-state chemical transformations e.g at interfaces between semiconductors and metallic contacts, and mechanical behavior \cite{McDowell2018}.  For in-situ TEM to be beneficial to superconductors, samples must be cooled to cryogenic temperatures and studied under the influence of magnetic fields.

Developed in the 1960s, liquid helium cooled stages have been used to study superconductors, solidified gases, and magnetic domains \cite{goodge_2020}, initially without the benefit of aberration-corrected systems with atomic-scale resolution.  More recently, cryogenic STEM with atomic scale resolution has been used to study quantum materials, including low-temperature spin states.  However, these studies have been limited to a single temperature that was above the boiling point of the choice cryogen (liquid helium or liquid nitrogen)\cite{goodge_2020}, due to thermal load, whereas variable temperature capabilities are requisite for probing phase transitions and the effects of thermal energy.  To this end, there has recently been a push to develop advanced sample holders with stable temperature control\cite{goodge_2020, HummingbirdSBIR}.  One of the most promising efforts is led by Hummingbird Precision Machine, a company that is developing a double-tilt, cryo-electrical biasing holder for TEMs that allows samples to be concurrently cooled to liquid helium temperatures and electrically biased, while undergoing atomic-scale structural imaging \cite{HummingbirdSBIR}.  Because of such industry involvement in the development and commercialization of cryogenic sample holders and, more generally, the rapid pace of in-situ TEM (e.g. the number of in-situ TEM papers doubled between 2010 and 2012\cite{Taheri2016}) we expect to see large advancements in this identified challenge over the next several years.

\subsubsection{Cross-sectional imaging of vortex structures}

In Sec.~\ref{sec:Introduction}, we discussed how competition between pinning forces, vortex elasticity, and current-induced forces results in complicated vortex structures, such as double-kinks, half-loops, and staircases. Typically, we may conjecture which structures have formed based on the microstructure and applied field orientation. Subsequent correlations are made between the presumed structures and the magnetization or transport results, which may be suggestive of a specific vortex phase. However, without direct proof of the structures, we cannot unequivocally correlate distinct excitations with specific vortex phases. For example, in a study of a NbSe$_2$ crystal containing columnar defects tilted 30$^\circ$ from the c-axis, magnetization results evinced glassy behavior when the field was aligned with the c-axis\cite{Eley2018}.  As these conditions are likely to produce vortex staircases, the question arose whether (and why) vortex staircases would create a vortex glass phase.

Direct imaging of vortex-defect interactions, in a way that captures the vortex structure overlaid on the atomic-scale structure, would enable unambiguous determination of the phases produced by specific vortex excitations. Accordingly, development of advanced microscopy techniques that can produce cross sectional images is another transformative opportunity in vortex physics.  In this section, we summarize common techniques for imaging superconducting vortices, detail their limitations, and describe the features of an advanced instrument that could accelerate progress in understanding and designing materials with predetermined vortex phases.


\begin{figure}
\includegraphics[width=1\linewidth]{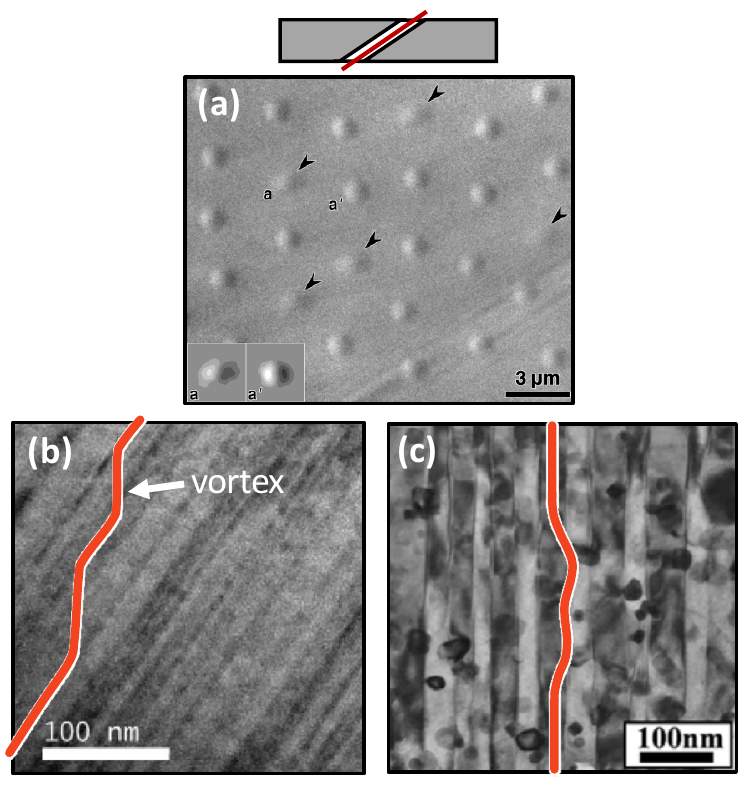}
\caption{Lorentz Transmission Electron Microscopy (LTEM) image of vortices in Bi$_2$Sr$_2$CaCu$_2$O$_{8+\delta}$, irradiated to induce columnar defects. Trapped vortices can be distinguished from free ones, based on their shape and contrast (lower contrast for vortices trapped in columnar defects). Plan-view LTEM images have provided useful information on vortex dynamics, though the 3D vortex structure within the bulk is hidden from view. Reprinted with permission from Ref.~[\onlinecite{Kamimura2002}].  Copyright 2002, \emph{The Physical Society of Japan}. TEM images of (b) a heavy-ion-irradiated NbSe$_2$ crystal\cite{Eley2018} and (c) a BaZrO$_3$-doped (Y$_{0.77}$Gd$_{0.23}$)Ba$_2$Cu$_3$O$_y$ film grown by Miura \textit{et al.} \cite{Miura2013k}. Permission to use TEM image in (c) granted by M. Miura.  Red line is a cartoon of how a vortex might wind through the disorder landscape. Advanced microscopy techniques designed to capture this structure is a transformative opportunity in vortex physics. \label{fig:vortexstructuresTEM}}
\end{figure}

Lorentz TEM (LTEM), which exploits the Aharonov-Bohm effect to capture magnetic contrast, was first used by Tonomura to image superconducting vortices~\cite{Tonomura2006} and has played a major role in identifying new materials that host exotic magnetic phases.  However, in LTEM, the objective lens serves the dual purpose of applying a field and observing the response of the specimen, and is therefore limited to plan-view imaging. That is, for out-of-plane magnetic fields applied to a thin film, the technique can only image magnetic contrast across the film’s surface---such that the vortex structure itself and interactions with defects within the bulk are out-of-view.

Building upon Tonomura's initial work, Hitachi\cite{Harada2008, Kawasaki2000} developed a unique, specialized system containing a multipole magnet that can apply fields up to \SI{50}{\milli\tesla} at various orientations with respect to the sample. Though this system still produces plan-view images, rather than cross-sectional images (revealing the full vortex structure), variations in the contrast of the imaged vortex section has provided remarkable evidence of vortex pinning and useful information on vortex-defect interactions.\cite{Kamimura2002, HARADA20131, PhysRevLett.88.237001, Tonomura2001}  For example, Fig.~\ref{fig:vortexstructuresTEM}(a) shows a LTEM image in which a vortex trapped within a columnar defect can be identified by its shape and contrast, compared to untrapped vortices.

The most promising technique for directly imaging vortex-defect interactions may be differential phase contrast microscopy (DPC)\cite{Lubk2015}. Conducted in a TEM, DPC is one of the best tools for quantitatively imaging nanoscale magnetic structures.  In a TEM, an illuminating electron beam is deflected by electromagnetic fields within a material. DPC microscopy leverages these deflections to directly image electric and magnetic fields within materials at atomic resolution\cite{Dekkers1974, Chapman1978}.  Consequently, scanning the beam (STEM) produces spatial maps of nanoscale magnetic field contrast to complement the atomic-scale structural information resolved by a transmitting beam.  Accordingly, STEM-DPC is an invaluable tool in nanomagnetism research, used to image magnetic domains\cite{Lee2017, Chen2018} and canted structures such as skyrmions\cite{McVitie2018, Matsumotoe1501280, Schneider2018}.  Most notably, it is one of few techniques that can unequivocally identify new magnetic phases and exotic magnetic quasiparticles in real-space. More generally, it can also image nanoscale electric fields\cite{Muller2014, Hachtel2018, Shibata2012, Shibata2017, Yucelen2018} in materials and devices.  To image vortex-defect structures in an STEM capable of DPC, the sample stage would need to be cryogenically cooled and the chamber should contain a magnet. Complications will include designing the system in a way in which the magnetic field does not significantly distort the beam.

\section{Summary and Outlook}
In this Perspective, we have highlighted the pivotal role that vortices play in superconductors and how improving our ability to control vortex dynamics will have an immediate impact on a range of applications. Herein we discussed major open questions in vortex physics, which include the following:

\begin{itemize}[noitemsep, leftmargin=*]
\item How do thermal and quantum vortex creep depend on material parameters and how can we efficiently consider creep in predictive simulations?

\item What is the highest attainable critical current $\Jc$?

\item How do we optimize vortex pinning in quantum circuits and controllably exploit vortices in certain schemes for topological computing? 

\item Given the multitude of variables that govern $\Jc$, what computational methods can improve the efficacy of the critical-current-by-design approach?

\item What is the relationship between $\Jc$ and point defect densities as well as vortex structures and vortex phases?
\end{itemize}

To answer these and other identified questions, we delineated five major categories of near-term transformative opportunities: The first involves applying recent advances in analytical and computational methods to model vortex creep, and performing more extensive experimental investigations into quantum creep.  Second, we discussed how critical currents higher than the current record of 30\% $J_d$ may be obtained by implementing a combination of core pinning and magnetic pinning. This is a promising route for dramatic advancements in large-scale applications---achieving higher currents densities enables smaller motors and generators as well as higher field magnets. Third, we noted that vortices do not only hamper large-scale applications, but also induce losses in nanoscale quantum circuits. Though shielding circuits has proven effective in minimizing vortex formation, quantum senors may require exposure to the environment, necessitating a better understanding of vortex dynamics in circuits. Furthermore, vortices are desirable for use in quantum information applications, in which case we must study how to manipulate single flux lines to implement braiding and entanglement of Majorana Bound States.  Fourth, the recent advent of high-performance computational tools to study vortex matter numerically has pushed us to the verge of predicting a superconductor's electrical transport properties based on the material and microstructure. However, the quest to automatically tailor a defect landscape for specific applications requires considering a fairly high-dimensional parameter space. To enable an effective mapping between simulations and experiments and manage the multitude of variables, we propose to apply self-adjusting machine learning algorithms that use neural networks. Fifth and finally, to accurately determine structure-property relationships, we need to experimentally measure and routinely consider point defect densities, which are challenging to determine. We therefore highlighted the prevailing microscopy techniques for point defect measurements, which include 4D-STEM and positron annihilation lifetime spectroscopy.

\begin{acknowledgments}
S.E.\ acknowledges support from the National Science Foundation DMR-1905909.
A.G.\ was supported by the U.S.\ Department of Energy, Office of Science, Basic Energy Sciences, Materials Sciences and Engineering Division.
R.W.\ acknowledges funding support from the Heidelberger Akademie der Wissenschaften through its WIN initiative (8. Teilprogramm).

\end{acknowledgments}

\section*{data availability}

The data that support the findings of this study are available from the corresponding author upon reasonable request.

\bibliography{perspective} 

\end{document}